\documentclass[copyright,creativecommons]{eptcs}
\providecommand{\event}{SYNT 2017} 

\usepackage[utf8]{inputenc}
\usepackage[mathscr]{euscript}
\usepackage{graphicx}
\usepackage{amssymb}
\usepackage{amsmath}
\usepackage{amsthm}

\usepackage{algorithm}
\usepackage{algorithmic}

\usepackage[usenames,dvipsnames]{color}

\usepackage[htt]{hyphenat}
\usepackage{url}
\usepackage{breakurl} 

\usepackage{hyperref}
\providecommand{\doi}[1]{\textsc{doi}: \href{http://dx.doi.org/#1}{\nolinkurl{#1}}}

\usepackage{xspace}
\usepackage{color}

\usepackage{booktabs} 

\usepackage{listings}
\definecolor{DarkGreen}{rgb}{0,0.3,0}
\definecolor{DarkBlue}{rgb}{0,0,0.7}
\definecolor{DarkRed}{rgb}{0.8,0,0}

\lstdefinelanguage{Spectra}[]{}{
  commentstyle=\color{DarkGreen}\itshape,
  keywordstyle=\color{blue}\bfseries,
  keywordstyle=[2]\color{DarkGreen}\bfseries,
  keywordstyle=[3]\color{DarkRed}\bfseries,
  morekeywords=[1]{G,GF,next,H,ONCE,boolean,pRespondsToS},
  morekeywords=[2]{define,in,out,module,gar,asm},
  morekeywords=[3]{sys,env},
  morecomment=[l]{//}
}

\usepackage{multicol}

\newenvironment{packed_itemize}{
\begin{itemize}
  \setlength{\itemsep}{1pt}
  \setlength{\parskip}{0pt}
  \setlength{\parsep}{0pt}
}{\end{itemize}}

\newenvironment{packed_itemize_two_col}{
\begin{multicols}{2}
\begin{itemize}
  \setlength{\itemsep}{1pt}
  \setlength{\parskip}{0pt}
  \setlength{\parsep}{0pt}
}{
\end{itemize}
\end{multicols}
}

\newcommand*\circlediamond{\vcenter{\hbox{\includegraphics[height=10pt]{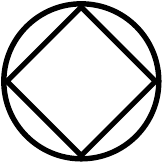}}}\xspace}
\newcommand*\circlebox{\vcenter{\hbox{\includegraphics[height=10pt]{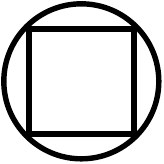}}}\xspace}

\newcommand{\ddmin}{DDMin\xspace}
\newcommand{\wsSpecs}{SYNTECH15\xspace}

\newcommand{\myTitle}{Performance Heuristics for GR(1) Synthesis\\ and
Related Algorithms}

\title{\myTitle}

\author{Elizabeth Firman \qquad Shahar Maoz \qquad Jan Oliver Ringert
\institute{School of Computer Science\\
Tel Aviv University, Israel}
}

\def\titlerunning{Performance Heuristics for GR(1) Synthesis and
Related Algorithms} 
\def\authorrunning{E. Firman, S. Maoz, \& J.O. Ringert}

\newcommand{\true}{{\small{\op{true}}}\xspace}
\newcommand{\false}{{\small{\op{false}}}\xspace}
\newcommand{\op}[1]{\textbf{\texttt{#1}}\xspace}

\begin{document}

\maketitle

\begin{abstract}
Reactive synthesis for the GR(1) fragment of LTL has been implemented and studied in many works.
In this workshop paper we present and evaluate a list of heuristics to
potentially reduce running times for GR(1) synthesis and related algorithms.
The list includes early detection of fixed-points and
unrealizability, fixed-point recycling, and heuristics for unrealizable core 
computations.
We evaluate the presented heuristics on SYNTECH15, a total of 78 specifications of 6
autonomous Lego robots, written by 3rd year undergraduate computer science
students in a project class we have taught, as well as on several benchmarks
from the literature.  The evaluation investigates not only the potential of 
the suggested heuristics to improve computation times, but also the difference
between existing benchmarks and the robot's specifications in terms of the
effectiveness of the heuristics.
\end{abstract}

\section{Introduction}

Reactive synthesis is an automated procedure to obtain a
correct-by-construction reactive system from its temporal logic
specification~\cite{PR89}.
Rather than manually constructing a system and using model checking to
verify its compliance with its specification, synthesis offers an
approach where a correct implementation of the system is automatically
obtained, if such an implementation exists.

GR(1) is a fragment of LTL, which has an efficient 
symbolic synthesis algorithm~\cite{BJP+12,PitermanPS06} and whose
expressive power covers most of the well-known LTL specification
patterns of Dwyer et al.~\cite{DAC99,MaozR15}.
GR(1) synthesis has been used and extended in different contexts and for
different application domains, including
robotics~\cite{Kress-GazitFP09}, scenario-based
specifications~\cite{MaozS12}, aspect languages~\cite{MaozS11},
event-based behavior models~\cite{DIppolitoBPU13}, and device
drivers~\cite{RyzhykW16}, to name a few.

In this workshop paper we present and investigate performance heuristics
for algorithms for GR(1) synthesis in case a specification is realizable
and Rabin(1) synthesis~\cite{KonighoferHB13,MaozS13AOSD} in case the
specification is unrealizable. For the case of unrealizability we also
investigate heuristics for speeding up the calculation of unrealizable
cores~\cite{CimattiRST08,KonighoferHB13}, i.e., minimal unrealizable
subsets that explain a cause of unrealizability.
For each heuristics we present (1) its rationale including the
source of the heuristics, if one exists, (2) how we implement it on top of
the basic algorithms, and (3) one example where the heuristics is
very effective and one example where it does not yield an improvement of
performance.

All heuristics we have developed and studied, satisfy three main
criteria. First, they are generic, i.e., they are not
optimized for a specific specification or family of specifications.
Second, they are all low risk heuristics, i.e., in the worst case they
may only have small negative effects on performance. Finally, they
are conservative, i.e., none of the heuristics changes the results
obtained from the algorithms.

We evaluate the presented heuristics on two sets of specifications. The
first set, SYNTECH15, consists of 78 specifications of 6 autonomous Lego
robots, written by 3rd year undergraduate computer science students in a
project class we have taught. The second set consists of specifications
for the ARM AMBA AHB Arbiter (AMBA) and a Generalized Buffer from an IBM
tutorial (GenBuf), which are the most popular GR(1) examples in
literature, used, e.g.,
in~\cite{BJP+12,CimattiRST08,KonighoferHB13,SchlaipferHB11}. Our
evaluation addresses the effectiveness of each of the heuristics
individually and together, and whether there exists a difference in
effectiveness with regard to different sets of specifications.

To the best of our knowledge, a comprehensive list of heuristics for
GR(1) and its systematic evaluation have not yet been published.

\section{Preliminaries}
\label{sec:preliminaries}

\paragraph{LTL and synthesis}

We repeat some of the standard definitions of linear temporal logic
(LTL), e.g., as found in~\cite{BJP+12}, a modal temporal logic with
modalities referring to time. LTL allows engineers to express properties
of computations of reactive systems. The syntax of LTL formulas is
typically defined over a set of atomic propositions $\mathit{AP}$ with
the future temporal operators \op{X} (next) and \op{U} (until).

The syntax of LTL formulas over $\mathit{AP}$
is $\varphi::=~p~|~\neg\varphi~|~
\varphi\vee\varphi~|~\op{X}\varphi~|~\varphi \op{U} \varphi$ for $p \in
\mathit{AP}$. The semantics of LTL formulas is defined over
computations. For $\Sigma = 2^\mathit{AP}$, a computation $u=u_0u_1..\in
\Sigma^\omega$ is a sequence where $u_i$ is the set of atomic propositions that hold at
the $i$-th position. For position $i$ we use $u,i\models \varphi$ to
denote that $\varphi$ holds at position $i$, inductively defined as: 

\begin{packed_itemize_two_col}
  \item $u,i\models p$ iff $p \in u_i$;
  \item $u,i\models \neg \phi$ iff $u, i \not\models \phi$;
  \item $u,i\models \varphi_1 \vee \varphi_2$ iff
  $u,i\models \varphi_1$ or $u,i\models \varphi_2$;
  \item $u,i\models \op{X}\varphi$ iff $u, i{+}1\models \varphi$;
  \item $u,i\models \varphi_1\op{U}\varphi_2$ iff $\exists k\geq i{:~}
  u,k\models\varphi_2$ and
  $\forall j, i\leq j < k{:~} u,j\models\varphi_1.$
\end{packed_itemize_two_col}

We denote $u,0\models \varphi$ by $u \models \varphi$. We use additional
LTL operators \op{F} (finally), \op{G} (globally), \op{ONCE} (at
least once in the past) and \op{H} (historically, i.e., always in the
past) defined as:

\begin{packed_itemize_two_col}
  \item $\op{F}\varphi := \true~\op{U}~\varphi$;
  \item $\op{G}\varphi := \neg\op{F}\neg\varphi$;
  \item $u,i\models \op{ONCE}\varphi$ iff $\exists 0 \leq k \leq i{:~}
  u,k\models \varphi$;
  \item $u,i\models \op{H}\varphi$ iff $\forall 0 \leq k \leq i{:~} u,k\models
  \varphi$.
\end{packed_itemize_two_col}

LTL formulas can be used as specifications of reactive systems where
atomic propositions are interpreted as environment (input) and system
(output) variables. An assignment to all variables is called a state.
Winning states are states from which the system can satisfy its
specification. A winning strategy for an LTL specification $\varphi$
prescribes the outputs of a system that from its winning states for all
environment choices lead to computations that satisfy $\varphi$. A
specification $\varphi$ is called realizable if a strategy exists such
that for all initial environment choices the initial states are winning
states. The goal of LTL synthesis is, given an LTL specification, to
find a strategy that realizes it, if one exists.

\paragraph{$\mu$-Calculus and Fixed-Points}

The modal $\mu$-calculus is a fixed-point logic~\cite{Kozen83}. It
extends modal logic with least ($\mu$) and greatest ($\nu$) fixed
points.
We use the $\mu$-calculus over the power set lattice of a finite set of
states $S$, i.e., the values of fixed-points are subsets of $S$. For
monotonic functions $\psi$ over this lattice and by the Knaster-Tarski
theorem the fixed points $\mu X.\psi(X)$ and $\nu Y. \psi(Y)$ are
uniquely defined and guaranteed to exist. The fixed-points can be
computed iteratively~\cite{2001automata} in at most $|S|$ iterations due
to monotonicity of $\psi$:

\begin{packed_itemize}
\item $\mu X.\psi(X)$: From $X_0:=\bot$ and $X_{i+1}:=\psi(X_i)$
obtain $\mu X.\psi(X) := X_{f}$ for $X_{f}=\psi(X_f)$  (note $f \leq
|S|$)
\item $\nu Y.\psi(Y)$: From $Y_0:=\top$ and $Y_{i+1}:=\psi(Y_i)$
obtain $\nu Y.\psi(Y) := Y_{f}$ for $Y_{f}=\psi(Y_f)$ (note $f \leq
|S|$)
\end{packed_itemize}

The fixed-point computation is linear in $|S|$. When states are
represented by a set of atomic propositions (or Boolean variables) $AP$
then $|S|=2^{|AP|}$, i.e., the number of iterations is exponential in
$AP$. Because the least (greatest) fixed-point is unique and $\psi$ is
monotonic we can safely start the iteration from under-approximations
(over-approximations). Good approximations can reduce the number of
iterations to reach the fixed-point.

\paragraph{GR(1) Synthesis}

GR(1) synthesis~\cite{BJP+12} handles a fragment of LTL where specifications
contain initial assumptions and guarantees over initial states, safety
assumptions and guarantees relating the current and next state, and justice
assumptions and guarantees requiring that an assertion holds infinitely many times during a
computation. A GR(1) synthesis problem consists of the following
elements~\cite{BJP+12}:
\begin{packed_itemize}
  \item $\mathcal{X}$ input variables controlled by the environment;
  \item $\mathcal{Y}$ output variables controlled by the system;
  \item $\theta^e$ assertion over $\mathcal{X}$ characterizing initial
  environment states;
  \item $\theta^s$ assertion over $\mathcal{X} \cup \mathcal{Y}$
  characterizing initial system states;
  \item $\rho^e(\mathcal{X} \cup \mathcal{Y}, \mathcal{X})$ transition
  relation of the environment;  
  \item $\rho^s(\mathcal{X} \cup \mathcal{Y}, \mathcal{X} \cup
  \mathcal{Y})$ transition relation of the system;  
  \item $J^e_{i \in 1..n}$ justice constraints of the environment to
  satisfy infinitely often;
  \item $J^s_{j \in 1..m}$ justice constraints of the system to
  satisfy infinitely often.
\end{packed_itemize}

GR(1) synthesis has the following notion of (strict)
realizability~\cite{BJP+12} defined by the LTL formula:
$$
\varphi^{sr} = (\theta^e \rightarrow \theta^s) \wedge (\theta^e
\rightarrow \op{G}((\op{H}\rho^e)\rightarrow \rho^s)) \wedge
(\theta^e
\wedge \op{G}\rho^e \rightarrow (\bigwedge_{i \in 1..n} \op{GF}
J_i^e \rightarrow \bigwedge_{j \in 1..m} \op{GF}
J_j^s)).$$

Specifications for GR(1) synthesis have to be expressible in the above structure
and thus do not cover the complete LTL. Efficient symbolic algorithms for GR(1) 
realizability checking and strategy synthesis for $\varphi^{sr}$ have been presented
in~\cite{BJP+12,PitermanPS06}.
The algorithm of Piterman et al.~\cite{PitermanPS06} computes winning
states for the system, i.e., states from which the system can ensure
satisfaction of $\varphi^{sr}$. 
We denote the states from which the system can force the environment to
visit a state in $R$ by $\circlediamond(R)$ defined as: 
$$
\circlediamond(R) = \{q \in 2^{\mathcal{X}\cup\mathcal{Y}}~|~ \forall x \in
2^\mathcal{X}:
\neg\rho^e(q,x) \vee \exists y\in 2^\mathcal{Y}:\\(\rho^s(q,\langle
x,y\rangle) \wedge \langle x,y\rangle \in R)\}.
$$
The system winning states are given by the following
formula using $\mu$-calculus notation:
\begin{equation}
W_{sys} = \nu Z.\bigcap_{j=1}^{m}\mu Y.\bigcup_{i=1}^{n} \nu X.\\
(J_j^s \cap \circlediamond(Z))\cup \circlediamond(Y) \cup (\neg J_i^e \cap
\circlediamond(X))\label{eqn:winSys}
\end{equation}
The algorithm from~\cite{BJP+12} for computing the set $W_{sys}$ is
shown in Alg.~\ref{alg:gr1}. Note that this algorithm already contains
some performance improvements over the naive evaluation of
Eqn.~(\ref{eqn:winSys}), e.g., the nested fixed-points $Y$ are not
computed independently for each $J^s_j$ and $Z$; instead
the value of $Z$ is updated before computing $J^s_{j+1}$.
Algorithm~\ref{alg:gr1} stores intermediate computation results in
arrays
\texttt{Z[]} (L.~\ref{alg:gr1:storeZ}), \texttt{Y[][]}
(L.~\ref{alg:gr1:storeY}), and \texttt{X[][][]}
(L.~\ref{alg:gr1:storeX}). This memory is used for strategy
construction~\cite{BJP+12}.

\begin{figure}
\begin{minipage}[t]{.49\textwidth}
\begin{algorithm}[H]
\caption{GR(1) game algorithm from~\cite{BJP+12} to compute system
winning states $Z$}
\label{alg:gr1}
\begin{algorithmic}[1]\footnotesize
  \STATE $Z = \true$
  \WHILE {not reached fixed-point of $Z$}
  \label{alg:gr1:fixZ}
  \FOR {$j=1$ \textbf{to} $|J^s|$}
    \label{alg:gr1:loopjs}
    \STATE $Y = \false$; $cy=0$
    \WHILE {not reached fixed-point of $Y$}
      \STATE $start = J^s_j \wedge \circlediamond Z \vee \circlediamond Y$
      \STATE $Y = \false$
      \FOR {$i=1$ \textbf{to} $|J^e|$}
        \STATE $X = Z$ \textit{ // better approx. than
        \true, see~\cite{BJP+12}}\label{alg:gr1:initX}
        \WHILE {not reached fixed-point of $X$}
          \STATE $X=start \vee (\neg J^e_i \wedge \circlediamond X)$
        \ENDWHILE
        \STATE {$Y = Y \vee X$}
        \STATE {\texttt{X[$j$][$i$][$cy$]}$\leftarrow
        X$}\label{alg:gr1:storeX}
      \ENDFOR
      \STATE {\texttt{Y[$j$][$cy$++]}$ \leftarrow Y$ }
      \label{alg:gr1:storeY}
    \ENDWHILE
    \STATE $Z=Y$
    \STATE {\texttt{Z[$j$]} = Y}\label{alg:gr1:storeZ}
    \label{alg:gr1:zUpdate}
    \label{alg:gr1:earlyUn}
  \ENDFOR
  \ENDWHILE
  \RETURN $Z$ 
\end{algorithmic}
\end{algorithm}
\end{minipage}
~
\begin{minipage}[t]{.49\textwidth}
\begin{algorithm}[H]
\caption{Rabin(1) game algorithm from~\cite{MaozS13AOSD,PnueliSZ10} to
compute environment winning states $Z$}
\label{alg:rabin}
\begin{algorithmic}[1]\footnotesize
  \STATE $Z = \false; cz=0$
  \WHILE {not reached fixed-point of $Z$}
  \FOR {$j=1$ \textbf{to} $|J^s|$}
    \label{alg:rabin:loopjs}
    \STATE $Y = \true$
    \WHILE {not reached fixed-point of $Y$}
      \STATE $start = \neg J^s_j \wedge \circlebox Y$
      \STATE $Y = \true$
      \FOR {$i=1$ \textbf{to} $|J^e|$}
        \STATE $pre = \circlebox Z \vee J^e_i \wedge start$
        \STATE $X = \false; cx=0$\label{alg:rabin:initX}
        \WHILE {not reached fixed-point of $X$}
          \STATE $X=pre \vee (\neg J^s_j \wedge \circlebox X)$ 
          \STATE {\texttt{X[$cz$][$i$][$cx$++]} $\leftarrow
          X$}\label{alg:rabin:storeX}
        \ENDWHILE
      \STATE $Y = Y \wedge X$
      \ENDFOR
    \ENDWHILE
    \STATE $Z=Z \vee Y$\label{alg:rabin:updateZ}
    \STATE {\texttt{Z[$cz$++]} $\leftarrow Y$}
    \label{alg:rabin:earlyUn}\label{alg:rabin:storeZ}
  \ENDFOR
  \ENDWHILE
  \RETURN $Z$
\end{algorithmic}
\end{algorithm}
\end{minipage}
\end{figure}

\paragraph{Unrealizability and Rabin(1) Game}

A specification $\varphi$ is unrealizable if there is a counter-strategy
in which the environment can force the system to violate at least one of
its guarantees while satisfying all the environment assumptions.
Maoz and Sa'ar~\cite{MaozS13AOSD} show how to compute the fixed-point
algorithm given by Könighofer et al.~\cite{KonighoferHB13} by playing a
generalized Rabin game with one acceptance pair (Rabin(1)
game\footnote{We use Rabin(1) to refer to the dual of GR(1) to avoid
confusion with ``Generalized Rabin(1) synthesis'' as defined by
Ehlers~\cite{Ehlers11}, where assumptions and guarantees are expressed
by generalized Rabin(1) conditions.}).
The algorithm computes the set of the winning states for the environment
by calculating cycles violating at least one justice guarantee $J^s_i$
while satisfying all justice assumptions $J^e_j$. Cycles can be left by
the system iff the environment can force it to a future cycle (ensures
termination) or to a safety guarantee violation.

We denote the states from which the environment can force the system to
visit a state in $R$ by $\circlebox(R)$ defined as: 
$$
\circlebox(R) = \{q \in 2^{\mathcal{X}\cup\mathcal{Y}}~|~ \exists x\in
2^\mathcal{X}:
\rho^e(q,x) \wedge \forall y\in
2^\mathcal{Y}: \\(\neg \rho^s(q,\langle x,y\rangle) \vee \langle
x,y\rangle \in R)\}.
$$
The set of environment wining states is given by the following formula
using $\mu$-calculus notation:
\begin{equation}
W_{env} = \mu Z.\bigcup_{j=1}^{m}\nu Y.\bigcap_{i=1}^{n} \mu X.\\
(\neg J_j^s \cup \circlebox(Z))\cap \circlebox(Y) \cap (J_i^e \cup
\circlebox(X))\label{eqn:winEnv}
\end{equation}

The algorithm from~\cite{MaozS13AOSD} (extended to handle $J^e$ as
implemented in JTLV~\cite{PnueliSZ10}) for computing the set $W_{env}$ is shown in
Alg.~\ref{alg:rabin}. Again, the algorithm already implements some
optimizations over the naive implementation of Eqn.~(\ref{eqn:winEnv}),
e.g., the early update of $Z$ in L.~\ref{alg:rabin:updateZ}.
Algorithm~\ref{alg:rabin} stores intermediate computation results in
arrays
\texttt{Z[]} (L.~\ref{alg:rabin:storeZ}) and \texttt{X[][][]}
(L.~\ref{alg:rabin:storeX}) for strategy
construction.

\paragraph{Delta Debugging (\ddmin)}
The Delta Debugging algorithm~\cite{Zeller99} (\ddmin) finds a locally
minimal subset of a set $E$ for a given monotonic criterion
\texttt{check}. We show the \ddmin algorithm in Alg.~\ref{alg:ddmin}.
The input of the algorithm are a set $E$ and the number $n$ of
partitions of $E$ to check. The algorithm starts with $n=2$ and refines
$E$ and $n$ in recursive calls according to different cases (L.~6,
L.~11, and L.~14).
The computation starts by partitioning $E$ into $n$ subsets and
evaluating \texttt{check} on each subset $part$ (L.~4) and its
complement (L.~10).
If \texttt{check} holds (L.~6 or L.~11), the search is continued
recursively on the subset $part$ (or its complement), until $part$ (or
its complement) has no subsets that satisfy \texttt{check}. If \texttt{check} neither
holds on any subset $part$ nor on the complements the algorithm
increases the granularity of the partitioning to $2n$ (L.~14) and
restarts.
 
One application of \ddmin is to find an
unrealizable core, a locally minimal subset of system guarantees for
which a specification is unrealizable.
To compute an unrealizable core the method \texttt{check} performs a
realizability check for the given subset $part$ of system guarantees.

\begin{algorithm}
\caption{Delta Debugging algorithm \ddmin from~\cite{Zeller99} as a
recursive method that minimizes a set of elements $E$ by partitioning
it into $n$ partitions (initial value $n=2$)}
\label{alg:ddmin}
\begin{minipage}[t]{.49\textwidth}
\begin{algorithmic}[1]\footnotesize
  \IF {$n > |E|$}
    \RETURN $E$
  \ENDIF
  \FOR {$part \in partition(E, n)$}\label{alg:ddmin:parts}
    \IF {\textbf{check}($part$)}\label{alg:ddmin:success}
      \RETURN \textbf{ddmin}($part$, 2)
    \ENDIF
  \ENDFOR
\end{algorithmic}
\end{minipage}
~
\begin{minipage}[t]{.49\textwidth}
\begin{algorithmic}[1]\footnotesize
  \setcounter{ALC@line}{8}
  \FOR {$part \in partition(E, n)$}\label{alg:ddmin:complements}
    \IF {\textbf{check}($E \setminus part$)}
      \RETURN \textbf{ddmin}($E \setminus part$, $n-1$)
    \ENDIF
  \ENDFOR
  \RETURN \textbf{ddmin}($E$, $min(|E|, 2n)$)\label{alg:ddmin:increase}
\end{algorithmic}
\end{minipage}
\end{algorithm}

\paragraph{Syntax in Examples}

Throughout the paper we present listings with example specifications
that describe GR(1) synthesis problems. We use the following syntax in
these specifications:

\begin{packed_itemize}
  \item $\mathcal{X}, \mathcal{Y}$: variables are either environment
  controlled ($\mathcal{X}$) and introduced by the keyword \texttt{env}
  or system controlled ($\mathcal{Y}$) and introduced by the keyword
  \texttt{sys}; variables have a type and a name, e.g., \op{sys
  boolean[4] button} declares a system variable of \op{boolean} array
  type of size 4 with the name \op{button}.
  \item $\theta^e, \rho^e, J^e$: assumptions are introduced by the
  keyword \op{asm}; initial assumptions, i.e., conjuncts of $\theta^e$,
  are propositional expressions over $\mathcal{X}$, safety assumptions,
  i.e., conjuncts of $\rho^e$, start with the temporal operator \op{G}
  and are propositional expressions over $\mathcal{X}$ and
  $\mathcal{Y}$ that may contain the operator \texttt{next} to refer to
  successor values of variables in $\mathcal{X}$, and justice
  assumptions, i.e., elements $J^e_i$, start with the temporal
  operators \op{GF} and are propositional expressions over
  $\mathcal{X}$ and $\mathcal{Y}$.
  \item $\theta^s, \rho^s, J^s$: guarantees are introduced by the
  keyword \op{gar}; guarantees are defined analogously to assumptions
  with the difference that $\theta^s$ may also refer to variables in
  $\mathcal{Y}$ and $\rho^s$ may apply the operator \op{next} also to
  variables in $\mathcal{Y}$.
\end{packed_itemize}

We denote propositional operators by standard symbols, i.e,
conjunction ($\wedge$) by \op{\&}, disjunction ($\vee$) by \op{|},
and negation ($\neg$) by \op{!}.

\section{Suggested Performance Heuristics}
\label{sec:optimizations}

We now present a list of heuristics for optimizing running times. The
first list applies to the GR(1) and Rabin(1) fixed-point algorithms
(Sect.~\ref{sec:heuristics:algorithms}).
The second list applies to computing unrealizable cores
(Sect.~\ref{sec:heuristics:core}). For each heuristics we present a
\textbf{rationale} including a source of the heuristics, the
\textbf{heuristics} and how we implemented it in
Alg.~\ref{alg:gr1}-\ref{alg:ddmin}, and two \textbf{examples} for
specifications where (1) the heuristics is effective and where (2) it
does not yield an improvement.

\subsection{GR(1) and Rabin(1) Fixed-Point Algorithm}
\label{sec:heuristics:algorithms}

\subsubsection{Early detection of fixed-point}
\label{efp}

\paragraph{Rationale.} The GR(1) game and the Rabin(1) game iterate over
the justice guarantees in the outermost fixed-point. Each iteration
refines the set of winning states based on the justice guarantee and the
calculated set from the previous iteration (\texttt{for}-loop in
Alg.~\ref{alg:gr1}, L.~\ref{alg:gr1:loopjs} and Alg.~\ref{alg:rabin},
L.~\ref{alg:rabin:loopjs}).
Computing a fixed-point for the same justice guarantee $J^s_j$ and the
same set $Z$ always yields the same set of winning states. We can exploit the
equality to detect if we will reach a fixed-point without completing the
\texttt{for}-loop, i.e., without computing the fixed-points for all
justice guarantees.
We found this heuristics implemented in the Rabin(1) game in
JTLV~\cite{PnueliSZ10}. We have not seen a similar implementation for
the GR(1) game.

\paragraph{Heuristics.} For each iteration of the justice guarantees
$J^s$ we save the resulting set of winning states for justice $J^s_j$
as \texttt{Z[$j$]} (Rabin(1), \texttt{Z[$cz$]}).
Starting in the second iteration of the outermost fixed-point we compare
for each justice $J^s_j$ the resulting $Z$ of its iteration to the
previously computed \texttt{Z[$j$]} (Rabin(1), \texttt{Z[$cz-|J^s|$]}). If
the sets are equal the algorithm reached a fixed-point with winning
states $Z$. The heuristics is correct since the next iteration of
justice $J^s_{j\oplus 1}$ will start from the set \texttt{Z[$j$]}
(Rabin(1), \texttt{Z[$cz-|J^s|$]}), which is the same set it started from
when it was previously computed. Hence, $\forall k>j:$ 
\texttt{Z[$k$]=Z[$j$]} (\texttt{Z[$cz-|J^s|$]=Z[$cz-|J^s| + k$]}), so by
definition we reached a fixed-point for $k=n$ (all justice guarantees).

\paragraph{Examples.}  

\begin{figure}[b]
\center{\textbf{Examples}: Early Detection of Fixed-Point}\\~\\
\lstset{language=Spectra}
\begin{minipage}{.49\textwidth}
\begin{lstlisting}[label=lst:EFPGood,caption={Heuristics very
effective}]
sys boolean[4] a;
gar G (a[0] = next(a[0])) &
      (a[1] = next(a[1])) &
      (a[2] = next(a[2])) &
      (a[3] = next(a[3]));
gar GF a[0] & a[1] & a[2] & a[3];
gar GF a[0];
gar GF a[1];
gar GF a[2];
\end{lstlisting}
\end{minipage}
\hfill
\begin{minipage}{.49\textwidth}
\begin{lstlisting}[label=lst:EFPBad,caption={Heuristics does not
yield improvement}]
sys boolean[4] a;
gar G (a[0] = next(a[0])) &
      (a[1] = next(a[1])) &
      (a[2] = next(a[2])) &
      (a[3] = next(a[3]));
gar GF a[0];
gar GF a[1];
gar GF a[2];
gar GF a[0] & a[1] & a[2] & a[3];
\end{lstlisting}
\end{minipage}
\end{figure}

Given the realizable GR(1) specification in
Listing~\ref{lst:EFPGood}, the standard GR(1) algorithm computes the set
of winning states in two iterations of the outer-most loop
(Alg.~\ref{alg:gr1}, L.~\ref{alg:gr1:fixZ}). The value of $Z$ becomes
$a[0] \wedge a[1] \wedge a[2] \wedge a[3]$ after the first step of the
loop over the justice guarantees $J^s$ (L.~\ref{alg:gr1:loopjs}). Early
fixed-point detection allows the algorithm to stop after checking
$J_1^s$ for the second time ($|J^s|+1$ executions of body of loop in
L.~\ref{alg:gr1:loopjs}) instead of going over all justice guarantees
again ($2\cdot|J^s|$ executions of body of loop in
L.~\ref{alg:gr1:loopjs}). For the similar specification in
Listing~\ref{lst:EFPBad} with a different order of justices early
fixed-point detection does not yield any improvement ($2\cdot|J^s|$
executions of body of loop in L.~\ref{alg:gr1:loopjs} are required) because the
last justice guarantee changed the fixed-point.

\subsubsection{Early detection of unrealizability}
\label{eun}

\paragraph{Rationale.} The GR(1) game and the Rabin(1) game compute
all winning states of the system and environment. When running GR(1)
synthesis or checking realizability we are interested whether there
exists a winning system output for all initial inputs from the
environment.
When running Rabin(1) synthesis or checking unrealizability we are
interested whether there is one initial environment input such that the
environment wins for all system outputs. Thus, in both cases it is not
necessary to compute all winning states, instead we can stop computation
once we can determine the outcome for the initial states.

\paragraph{Heuristics.} 
The outermost fixed-point in the GR(1) game is a greatest
fixed-point. The game starts from the set of all states and refines it
to the winning states. Thus, after the computation of the winning states
for a justice guarantee we check whether the system still wins from all
initial inputs. We implemented this check in Alg.~\ref{alg:gr1} after
L.~\ref{alg:gr1:earlyUn}. If the system loses for at least one initial
environment input we stop the computation of winning states.

The outermost fixed-point in the Rabin(1) game is a least
fixed-point. The game starts from an empty set of states and extends it
to the winning states. Thus, after the computation of the winning states
for a justice guarantee we check whether the environment now wins from
some initial input. We implemented this check in Alg.~\ref{alg:rabin}
after L.~\ref{alg:rabin:earlyUn}. If the environment wins for at least
one initial input we stop the computation of winning states.

\paragraph{Examples.}  
Given the unrealizable GR(1) specification in
Listing~\ref{lst:EDoUGood}, the standard GR(1) algorithm computes the
system winning states starting with all possible values of \texttt{c}.
In every iteration of the $Z$ fixed-point (see Alg.~\ref{alg:gr1},
L.~\ref{alg:gr1:fixZ}) two states are removed (the states with largest
uneven and even value of \texttt{c}).
For an integer domain \texttt{0..}$n$ ($n$=10000 in
Listing~\ref{lst:EDoUGood}) the GR(1) algorithm will compute $n/2$
justice guarantee iterations. Our heuristics will compute only 2 justice
guarantee iterations for the example shown in
Listing~\ref{lst:EDoUGood}. The heuristics will not yield an improvement
over the regular GR(1) implementation for the example shown in
Listing~\ref{lst:EDoUBad}. Here the losing initial state is only
detected in iteration $n/2$.

The same examples are also effective and non-effective examples for the
Rabin(1) game algorithm.

\begin{figure}
\center{\textbf{Examples}: Early Detection of Unrealizability}\\~\\
\lstset{language=Spectra}
\begin{minipage}{.49\textwidth}
\begin{lstlisting}[label=lst:EDoUGood,caption={Heuristics very
effective}]
sys Int(0..10000) c;
gar c=10000; 
gar G next(c)=c+1; 
gar GF (c mod 2 = 1);
\end{lstlisting}
\end{minipage}
\hfill
\begin{minipage}{.49\textwidth}
\begin{lstlisting}[label=lst:EDoUBad,caption={Heuristics does not
yield improvement}]
sys Int(0..10000) c;
gar c=0; // only difference
gar G next(c)=c+1; 
gar GF (c mod 2 = 1);
\end{lstlisting}
\end{minipage}
\end{figure}

\subsubsection{Fixed-point recycling}
\label{fpr}
\paragraph{Rationale.} 
The GR(1) game and the Rabin(1) game are solved by computing nested
fixed-points of  
monotonic functions (see Eqn.~(\ref{eqn:winSys}) and
Eqn.~(\ref{eqn:winEnv})). 
The time complexity of a straightforward implementation of the
fixed-point computation is cubic in the state space and can be reduced
to quadratic time~\cite{BrowneCJLM97}, as mentioned in~\cite{BJP+12}.
This method can also be applied to the Rabin(1) game. Interestingly,
although fixed-point recycling is used to obtain quadratic
instead of cubic time complexity of the GR(1) algorithm \cite{BJP+12}, to
the best of our knowledge no GR(1) tool has implemented it
following~\cite{BrowneCJLM97} and it has never been systematically
evaluated.

\paragraph{Heuristics.} 

Fixed-points are usually computed by fixed-point iteration starting from
$\bot$ (least fixed-points) or $\top$ (greatest fixed-points) until a
fixed point is reached. The same principle works for the evaluation of
nested fixed-points where for each iteration step of the outer
fixed-point, the inner fixed-point is computed from scratch.
The main idea of~\cite{BrowneCJLM97} is to exploit the monotonicity of
fixed-point computations and start nested fixed-point calculations from
approximations computed in earlier nested computations. Consider the
formula $\mu Z. \nu Y. \mu X.\psi(Z, Y, X)$, iteration $k+1$ of
$Z$, and iteration $l$ of $Y$: due to monotonicity $Z_k \subseteq
Z_{k+1}$ and $Y_{l}^{\mathit{of}~Z_{k}} \subseteq
Y_{l}^{\mathit{of}~Z_{k+1}}$. Thus, the fixed-point $X$ for
$Z_{k}$ and $Y_{l}^{\mathit{of}~Z_{k}}$ is an
under-approximation of the fixed-point $X$ for $Z_{k+1}$ and
$Y_{l}^{\mathit{of}~Z_{k+1}}$ (see~\cite{BrowneCJLM97} for more
details).

In both, the GR(1) algorithm and the Rabin(1) algorithm, the fixed-point
computations also depend on justice assumptions $J^e_i$ and justice
guarantees $J^s_j$. This dependence does not interfere with monotonicity
of the computation. However, the algorithms compute $|J^e|\cdot|J^s|$
values of the fixed-point $X$ for each iteration of $Y$ (stored in array
\texttt{X[][][]} in Alg.~\ref{alg:gr1}, L.~\ref{alg:gr1:storeX}).

We implemented this heuristics in the GR(1) game Alg.~\ref{alg:gr1} with
a modified start value for the fixed-point computation of $X$ in
L.~\ref{alg:gr1:initX}. Unless the algorithm computes the first
iteration of $Z$ the value of $X$ is set to the previously computed
result for the same justice assumption $J^e_i$ and justice guarantee
$J^s_j$ and same iteration $cy$ of $Y$, i.e., $X$ is set to memory cell
\texttt{X[$j$][$i$][$cy$]} intersected with $Z$. This value is an
over-approximation of the greatest fixed-point $X$ and its computation likely terminates
after fewer iterations.

Similarly, we implemented the fixed-point recycling heuristics in the
Rabin(1) game Alg.~\ref{alg:rabin} with a modified start value for the
fixed-point computation of $X$ in L.~\ref{alg:rabin:initX}. Unless the
algorithm computes the first iteration of $Z$ the value of $X$ is set to
the previously computed result for the same justice assumption $J^e_i$
and justice guarantee $J^s_j$ for the same iteration of $Y$. This value
is an under-approximation of the least fixed-point $X$ and its
computation likely terminates after fewer iterations. Note that in
Alg.~\ref{alg:rabin} the fixed point value of $X$ is only stored for the
last iteration of $Y$ (L.~\ref{alg:rabin:storeX}). We had to change the
implementation to store $X$ for all iterations of $Y$ to use fixed-point
recycling as described in~\cite{BrowneCJLM97}.

It is important to note that this heuristics
changes the worst-case running time of both algorithms from
$O(|J^e|\cdot|J^s|\cdot|N|^3)$ to
$O(|J^e|\cdot|J^s|\cdot|N|^2)$~\cite{BJP+12,BrowneCJLM97}.

\begin{figure}
\center{\textbf{Examples}: Fixed-Point Recycling}\\~\\
\lstset{language=Spectra}
\begin{minipage}{.49\textwidth}
\begin{lstlisting}[label=lst:FPRGood,caption={Heuristics very
effective}]
sys Int(0..10000) c;
sys boolean two;
gar G two; // force two Z-iterations
gar G (next(c) = c+1) | 
      (c=10000 & next(c) = 0);
gar GF c = 0;
asm GF c = 10000;
\end{lstlisting}
\end{minipage}
\hfill
\begin{minipage}{.49\textwidth}
\begin{lstlisting}[label=lst:FPRBad,caption={Heuristics does not
yield improvement}]
sys Int(0..10000) c;
sys boolean two;
gar G two; // force two Z-iterations
gar G (next(c) = c+1) | 
      (c=10000 & next(c) = 0);
gar GF c = 0;
asm GF c = 0; // only difference
\end{lstlisting}
\end{minipage}
\end{figure}

\paragraph{Examples.}  Consider the realizable GR(1) specification in
Listing~\ref{lst:FPRGood}. The variable \texttt{c} models a counter from 0
to 10,000 that increases and resets to 0 when reaching 10,000. The second
variable \texttt{two} serves only the purpose of ensuring two iterations of the $Z$
fixed-point (recycling cannot happen in the first iteration). In the
first iteration of $Z$ and $Y$ the nested computation of the $X$ fixed-point
requires 10,000 iterations (in each iteration losing one state to end
with \texttt{two \& x=0}). In the second $Z$ and first $Y$ iteration the
same computation repeats. Here, the fixed-point recycling heuristics
starts from \texttt{two \& x=0} and finishes after one iteration
instead of additional 10,000.
It is important to note that on the same specification without variable
\texttt{two} the heuristics would not yield an improvement because a
single $Z$ iteration is enough to detect that all states are winning
states. As another example for no improvement, consider the slightly
modified specification from Listing~\ref{lst:FPRBad}. Here the
single justice guarantee and justice assumption coincide and 
each nested computation of the $X$ fixed-point requires two iterations
with and without recycling.

%
%
%
%

\subsection{Unrealizable Core Calculation}
\label{sec:heuristics:core}

\subsubsection{Contained sets}
\label{sec:ddmin:containedSets}

\paragraph{Rationale.} 

The delta debugging algorithm \ddmin shown
in Alg.~\ref{alg:ddmin} might check subsets of guarantees which are
contained in previously checked realizable subsets
(e.g., after increasing the number of partitions to $2n$ when
all other checks failed). In these cases we don't have to execute the
costly realizability check: a subset $part$ of a realizable set
$E$ (failure of \texttt{check($E$)}) is also realizable.

This heuristics was mentioned in~\cite{ZellerH02} and also implemented
for unrealizable core calculation in~\cite{KonighoferHB13}.

\paragraph{Heuristics.} We extend the generic \ddmin algorithm shown in
Alg.~\ref{alg:ddmin}. Before checking a candidate set $E'$, i.e.,
executing \texttt{check($E'$)}, we look up whether $E'$ is a subset of
any previously checked set $E$ with negative evaluation of
\texttt{check($E$)}.

\paragraph{Examples.} 

\begin{figure}
\center{\textbf{Examples}: Contained Sets in \ddmin}\\~\\
\lstset{language=Spectra}
\begin{minipage}{.49\textwidth}
\begin{lstlisting}[label=lst:DDSGood,caption={Heuristics very
effective}]
sys boolean x;
gar g1: x;
gar g2: G TRUE; 
gar g3: G TRUE;
gar g4: G !x;
\end{lstlisting}
\end{minipage}
\hfill
\begin{minipage}{.49\textwidth}
\begin{lstlisting}[label=lst:DDSBad,caption={Heuristics does not
yield improvement}]
sys boolean x;
gar g1: FALSE;
gar g2: G TRUE; 
gar g3: G TRUE;
gar g4: G TRUE; 
\end{lstlisting}
\end{minipage}
\end{figure}

Given the unrealizable GR(1) specification in Listing~\ref{lst:DDSGood},
the computation of an unrealizable core based on \ddmin from
Alg.~\ref{alg:ddmin} calls the method \texttt{check} with the following
subsets of guarantees (positive results of \texttt{check} are
\underline{underlined}):
$\{\texttt{g1},\texttt{g2}\}$ (L.~5, $n=2$),
$\{\texttt{g3},\texttt{g4}\}$ (L.~5, $n=2$), 
$\{\texttt{g3},\texttt{g4}\}^*$ (L.~10, $n=2$), 
$\{\texttt{g1},\texttt{g2}\}^*$ (L.~10, $n=2$), 
$\{\texttt{g1}\}^*$ (L.~5, $n=4$),
$\{\texttt{g2}\}^*$ (L.~5, $n=4$),
$\{\texttt{g3}\}^*$ (L.~5, $n=4$),
$\{\texttt{g4}\}^*$ (L.~5, $n=4$),
$\{\texttt{g2},\texttt{g3},\texttt{g4}\}$ (L.~10, $n=4$),
\underline{$\{\texttt{g1},\texttt{g3},\texttt{g4}\}$} (L.~10, $n=4$),
$\{\texttt{g1}\}^*$ (L.~5, $n=3$),
$\{\texttt{g3}\}^*$ (L.~5, $n=3$),
$\{\texttt{g4}\}^*$ (L.~5, $n=3$),
$\{\texttt{g3},\texttt{g4}\}^*$ (L.~10, $n=3$),
\underline{$\{\texttt{g1},\texttt{g4}\}$} (L.~10, $n=3$),
$\{\texttt{g1}\}^*$ (L.~5, $n=2$),
$\{\texttt{g4}\}^*$ (L.~5, $n=2$),
$\{\texttt{g4}\}^*$ (L.~10, $n=2$), and
$\{\texttt{g1}\}^*$ (L.~10, $n=2$). 
Out of these 19 calls to \texttt{check} the described heuristics will
avoid running the realizability check in the 13 cases marked with a
star ($^*$). Given the similar unrealizable specification in
Listing~\ref{lst:DDSBad}, the described heuristics does not yield any improvement. The method \texttt{check} is
never invoked on a subset that it failed on. It is invoked on: 
\underline{$\{\texttt{g1},\texttt{g2}\}$} (L.~5, $n=2$) and
\underline{$\{\texttt{g1}\}$} (L.~5, $n=2$).

\subsubsection{Incremental GR(1) for similar candidates}
\label{sec:ddmin:incremental}

\paragraph{Rationale.}
Due to the nature of the \ddmin algorithm (Alg.~\ref{alg:ddmin}), there are
multiple calls to check realizability of subsets of guarantees. Some of
the subsets share elements. We can try to reuse computation results from
previous calls to \texttt{check} for related subsets of guarantees to
speed up the computation of fixed-points, both in Rabin(1) and GR(1) games.

\paragraph{Heuristics.}
The main idea is to reuse results of previous computations of the GR(1)
game (Alg.~\ref{alg:gr1}) or the Rabin(1) game (Alg.~\ref{alg:rabin}). 
We identified three cases in \ddmin (Alg.~\ref{alg:ddmin}). In each
case we use different methods to reuse the computations from previous rounds.

Case 1: An unrealizable subset $parent$ was found (the set $part$ in
Alg.~\ref{alg:ddmin}, L.~\ref{alg:ddmin:success}) and \ddmin descends to
perform the search on subsets of $parent$, starting with $n=2$.
We examine the differences between $parent$ and its current subset of
guarantees to check. We have the following scenarios:

1. Only initial guarantees were removed from $parent$: In both the 
GR(1) and Rabin(1) games we can reuse the winning states ($Z$ in Alg.~\ref{alg:gr1} and
Alg.~\ref{alg:rabin}) that were computed for $parent$, and preform
only a simple check for realizability.
For GR(1) we check if the system can win from all its initial states.
For Rabin(1) we check if the environment can win for some of its initial
state.

2. Only safety guarantees were removed from $parent$: Since there are
less constraints the attractors $Y$ are larger, hence the set of winning
states $Z$ can be larger. In GR(1) we compute $Z$ using greatest
fixed-point, so we cannot reuse the previously computed $Z_{prev}$ to
initialize $Z$. However, $Z_{prev}$ is equivalent to the values $Y$
stored as \texttt{Z[$j$]}in Alg.~\ref{alg:gr1}, L.~\ref{alg:gr1:storeZ}
in the last fixed-point iteration of $Z$. Thus, $Z_{prev}$ is a safe
under-approximation of the least fixed-point $Y$ and we change the
initialization of $Y$ in line 4 to $Y=Z_{prev}$.

3. Only justice guarantees were removed from $parent$: We can
reuse all information of the previous computation up to the first
removed justice guarantee. We reuse the memory $\texttt{Z}_{prev}$,
$\texttt{Y}_{prev}$, and $\texttt{X}_{prev}$ from the first iteration
of $Z$ on $parent$ up to the first removed justice guarantee. Then
we continue the computation.

Case 2: All subsets $part$ of $parent$ are realizable and \ddmin
continues with complements in Alg.~\ref{alg:ddmin},
L.~\ref{alg:ddmin:complements}: In this case and for $n>2$ the
candidates $E \setminus part$ contain previously checked and realizable
candidates. Our main observation is that the system winning states for
guarantees $E \setminus part$ cannot be more than for any of its
subsets. 
We can check
realizability of a GR(1) game by initializing its greatest fixed-point
$Z$ to the intersection of system winning states $Z_{prev}$ of
previously computed subsets.
Alternatively, we can check realizability with a Rabin(1) game by
initializing its least fixed point $Z$ to the union of
environment winning states $Z_{prev}$ of previously computed subsets.

Case 3: All subsets and complements are realizable and \ddmin increases
search granularity in Alg.~\ref{alg:ddmin},
L.~\ref{alg:ddmin:increase}: For the new run Case 1 applies (with the
previous parent) and Case 2 applies when checking complements of the
sets with higher granularity.

\begin{figure}
\center{\textbf{Examples}: Incremental GR(1) in \ddmin}\\~\\
\lstset{language=Spectra}
\begin{minipage}{.49\textwidth}
\begin{lstlisting}[label=lst:IncGood,caption={Heuristics very
effective}]
sys boolean x;
sys boolean y;
gar g1: G !y;
gar g2: G !x;
gar g3: GF !y;
gar g4: G x;
\end{lstlisting}
\end{minipage}
\hfill
\begin{minipage}{.49\textwidth}
\begin{lstlisting}[label=lst:IncBad,caption={Heuristics does not
yield improvement}]
sys boolean x;
sys boolean y;
gar g1: G !y;
gar g2: G next(!x); // changed
gar g3: GF !y;
gar g4: GF x; // changed
\end{lstlisting}
\end{minipage}
\end{figure}

\paragraph{Examples.} The specification in Listing~\ref{lst:IncGood} is
unrealizable because the system cannot satisfy \texttt{g2} and
\texttt{g4} together. The first set that includes both guarantees
in a check of \ddmin (Alg.~\ref{alg:ddmin},
L.~\ref{alg:ddmin:complements}) is $\{\texttt{g2}, \texttt{g3},
\texttt{g4}\}$. Previously computed winning states are
states with \texttt{x=\true} for $\{\texttt{g2}\}$ and \texttt{x=\false}
for $\{\texttt{g3, g4}\}$. Their intersection is empty and determines
that $\{\texttt{g2}, \texttt{g3}, \texttt{g4}\}$ is unrealizable without
even playing a game. The second specification in
Listing~\ref{lst:IncBad} is very similar. Again the reason for
unrealizability are guarantees \texttt{g2} and
\texttt{g4}. However, at the same \ddmin step as before the previously
computed winning states for subsets of $\{\texttt{g2}, \texttt{g3},
\texttt{g4}\}$ are all states for $\{\texttt{g2}\}$ and all states
for $\{\texttt{g3, g4}\}$. The intersection of these winning states is
still the set of all states. In this case our incremental heuristics
does not yield improvement.

\subsubsection{GR(1) game vs. Rabin(1) game}

\paragraph{Rationale.} GR(1) games and Rabin(1) games are determined:
each game is either unrealizable for the system player or unrealizable
for the environment player. To check for unrealizability, it is thus
equally possible to play the Rabin(1) game or GR(1) game. 

The implementations of Könighofer et al.~\cite{KonighoferHB13} and
Cimatti et al.~\cite{CimattiRST08} use the GR(1) game for checking
realizability during unrealizable core computation.

\paragraph{Heuristics.} We replace the implementation of
\texttt{check}. Instead of playing the GR(1) game we play the Rabin(1) game  
and negate the result.

\paragraph{Examples.} The specification in Listing~\ref{lst:RabinGood}
is unrealizable because the environment can force the system to a
deadlock state: the states $\texttt{x = 127}$ have no successor for
environment input \texttt{y = \true}.
Both the Rabin(1) game and the GR(1) game require $O(n)$ (here $n=127$)
$Z$ iterations to compute the $Z$ fixed-point. Each $Z$ iteration
requires two $Y$ iterations. In the Rabin(1) game, each $Y$ iteration
requires two $X$ iterations. However, in the GR(1) game\footnote{For
this example we assume initialization of $X=\true$ in
Alg.~\ref{alg:gr1}, L.~\ref{alg:gr1:initX} instead of $Z$.
Note that the optimization of $X=Z$ from~\cite{BJP+12}, that we used in
all our experiments as base case, achieves fewer $X$ iterations.}
another $O(n)$ $X$ iterations are required for each $Y$ iteration.
For the similar specification in Listing~\ref{lst:RabinBad} the numbers
of fixed-point iterations of the Rabin(1) game are the same and here
also coincide with the number of iterations of the GR(1) game and the
heuristics does not contribute.

\begin{figure}
\center{\textbf{Examples}: GR(1) game vs. Rabin(1) game}\\~\\
\lstset{language=Spectra}
\begin{minipage}{.49\textwidth}
\begin{lstlisting}[label=lst:RabinGood,caption={Heuristics very
effective}]
env boolean y;
sys Int(0..127) x;
asm GF !y;
gar G y -> next(x)=x+1;
\end{lstlisting}
\end{minipage}
\hfill
\begin{minipage}{.49\textwidth}
\begin{lstlisting}[label=lst:RabinBad,caption={Heuristics does not
yield improvement}]
env boolean y;
sys Int(0..127) x;

gar G y -> next(x)=x+1;
\end{lstlisting}
\end{minipage}
\end{figure}

\section{Evaluation}

Our evaluation is divided into two parts following the  division of
heuristics into performance heuristics for the GR(1) and the Rabin(1)
algorithm from Sect.~\ref{sec:heuristics:algorithms} and performance
heuristics for calculating unrealizable cores from
Sect.~\ref{sec:heuristics:core}.
For both, we address the following two research
questions:
\begin{description}
\item[RQ1] What is the effectiveness of each of the heuristics
individually and together?
\item[RQ2] Is there a difference in effectiveness with regard to
different sets of specifications?
\end{description}

\subsection{Procedure}
\label{sec:procedure}
We used the GR(1) game and Rabin(1) game implementations shown in
Alg.~\ref{alg:gr1} and Alg.~\ref{alg:rabin} as reference (recall that
these algorithms already contain performance improvements over
naive implementations following the fixed-point formulation, see
Sect.~\ref{sec:preliminaries}).
We have implemented these two algorithms and all our suggested 
heuristics in C using CUDD 3.0~\cite{CUDD}. We measure
running-times in nanoseconds using C APIs. Our implementation starts
with the BDD variable order as it appears in the specification. We use
the default dynamic variable reordering of CUDD.

We have executed each realizability check for every specification
50 times (see Sect.~\ref{sec:threats}). We aggregated the 50 runs of
each specification as a median. The ratios we report are ratios of medians of each heuristics
compared to a base case (original implementations of algorithms as shown in
Alg.~\ref{alg:gr1}-\ref{alg:ddmin}) for the same specification.

\subsection{Evaluation Materials}
\label{sec:materials}

Only few GR(1) specifications are available and these were usually created by
authors of synthesis algorithms or extensions thereof. 

For the purpose of evaluation, we have used specifications created by
3rd year CS students in a workshop project class that we have taught.
Over the course of a semester, the students have created specifications
for the following systems, which they actually built and run:
ColorSort -- a robot sorting Lego pieces by color; Elevator -- an
elevator servicing different floors; Humanoid -- a mobile robot of
humanoid shape; PCar -- a self parking car; Gyro -- a robot with
self-balancing capabilities; and SelfParkingCar - a second version of a
self parking car. We call this set of specifications \wsSpecs.

The specifications were \textit{not} created specifically for the
evaluation in our paper but as part of the ordinary work of the students
in the workshop class. During their work spanning one semester, the
students have committed many versions of their specifications to the repository. In total, we
have collected 78 specifications. We consider these GR(1)
specifications to be the most realistic and relevant examples one could
find for the purpose of evaluating our work.

In addition to the specifications created by the students, we considered
the ARM AMBA AHB Arbiter (AMBA) and a Generalized Buffer from an IBM
tutorial (GenBuf), which are the most popular GR(1) examples in
literature, used, e.g.,
in~\cite{BJP+12,CimattiRST08,KonighoferHB13,SchlaipferHB11}. We included
5 different sizes of AMBA (1 to 5 masters) and 5 different sizes of
GenBuf (5 to 40 requests), each in its original version plus the 3
variants of unrealizability described in~\cite{CimattiRST08} (justice
assumption removed, justice guarantee added, and safety guarantee
added). We have thus run our experiments also on 20 AMBA and 20 GenBuf
specifications.

All specifications used in our evaluation, the raw data recorded from
all runs, and the program to reproduce our experiments are available
from~\cite{optimizationWebsite}.

\subsection{Evaluation Results}
\label{sect:evaluation}

We now present aggregated data from all runs on all specifications
with different heuristics and their combination. We decided to present
for all experiments minimum, maximum, and quartiles of ratios.

\subsubsection{Results for GR(1)/Rabin(1) Fixed-Point Algorithms}
We present the ratios of running times for heuristics from
Sect.~\ref{sec:heuristics:algorithms} separately for realizable and
unrealizable specifications from the set \wsSpecs and AMBA and GenBuf.
The different heuristics are abbreviated as follows:
\emph{efp} is the early fixed point detection from Sect.~\ref{efp},
\emph{eun} is the early unrealizability detection from Sect.~\ref{eun},
and \emph{fpr} is the fixed-point recycling from Sect.~\ref{fpr}.
By \emph{all} we refer to the use of all heuristics
together. All results are rounded to two decimals.
Tbl.~\ref{tbl:gr1_real} shows the ratios of running times for 61
realizable \wsSpecs specifications (top) and for 10 realizable AMBA and GenBuf
specifications (bottom). 
Tbl.~\ref{tbl:gr1_unreal} shows the ratios of running times for 17
unrealizable \wsSpecs specifications (top) and for 30 unrealizable AMBA
and GenBuf specifications (bottom).
All tables show first ratios of running times for the GR(1) algorithm,
then ratios for the Rabin(1) algorithm, and finally a comparison between the
Rabin(1) and GR(1) algorithms.

\begin{table}[t]\small
\centering
\rotatebox[origin=c]{90}{\wsSpecs\hspace{1em}}
\rotatebox[origin=c]{90}{realizable\hspace{1em}}
\hspace{1em}
\begin{tabular}{l | r r r r | r r r r | r r}
~~ & \multicolumn{4}{|c|}{GR(1) algorithm} & \multicolumn{4}{|c|}{Rabin(1)
algorithm} & \multicolumn{2}{|c}{Rabin(1) / GR(1)}\\
\hline
Quartile & efp & eun & fpr & all & efp & eun & fpr & all & orig & all\\
\hline
MIN & 0.61 & 0.94 & 0.6 & 0.53 & 0.59 & 0.92 & 0.6 & 0.52 & 0.52 & 0.46\\
$Q_1$ & 0.95 & 1 & 0.93 & 0.9 & 0.94 & 0.99 & 0.94 & 0.9 & 0.84 & 0.85\\
$Q_2$ & 0.99 & 1 & 0.96 & 0.95 & 0.98 & 1 & 0.96 & 0.95 & 0.91 & 0.91\\
$Q_3$ & 1 & 1.02 & 1 & 0.98 & 1 & 1 & 0.99 & 0.99 & 0.96 & 0.97 \\
MAX & 1.09 & 1.11 & 1.1 & 1.12 & 1.04 & 1.08 & 1.04 & 1.05 & 1.29 & 1.34\\

\end{tabular}\\
\vspace{.5em}
\centering
\rotatebox[origin=c]{90}{AMBA/GenBuf\hspace{1em}}
\rotatebox[origin=c]{90}{realizable\hspace{1em}}
\hspace{1em}
\begin{tabular}{l | r r r r | r r r r | r r}
~~ & \multicolumn{4}{|c|}{GR(1) algorithm} & \multicolumn{4}{|c|}{Rabin(1) algorithm} &
\multicolumn{2}{|c}{Rabin(1) / GR(1)}\\
\hline
Quartile & efp & eun & fpr & all & efp & eun & fpr & all & orig & all\\
\hline
MIN & 0.83 & 0.97 & 0.74 & 0.66 & 0.84 & 0.99 & 0.6 & 0.58 & 0.83 & 0.82\\
$Q_1$ & 0.93 & 0.99 & 0.83 & 0.82 & 0.91 & 1 & 0.86 & 0.84 & 0.88 & 0.88\\
$Q_2$ & 0.99 & 1 & 0.92 & 0.9 & 0.99 & 1 & 0.92 & 0.91 & 0.92 & 0.92 \\
$Q_3$ & 1 & 1 & 0.95 & 0.94 & 1 & 1 & 0.96 & 0.96 & 0.95 & 0.94\\
MAX & 1 & 1.01 & 0.96 & 0.96 & 1.01 & 1.02 & 0.99 & 0.97 & 1.05 & 1.04\\

\end{tabular}
\caption{\small Ratios of the heuristics to the original GR(1)
and Rabin(1) running times for realizable specifications.}
\label{tbl:gr1_real}
\end{table}

\begin{table}[t]\small
\centering
\rotatebox[origin=c]{90}{\wsSpecs\hspace{1em}}
\rotatebox[origin=c]{90}{unrealizable\hspace{1em}}
\hspace{1em}
\begin{tabular}{l | r r r r | r r r r | r r}
~~ & \multicolumn{4}{|c|}{GR(1) algorithm} & \multicolumn{4}{|c|}{Rabin(1) algorithm} &
\multicolumn{2}{|c}{Rabin(1) / GR(1)}\\
\hline
Quartile & efp & eun & fpr & all & efp & eun & fpr & all & orig & all\\
\hline
MIN & 0.94 & 0.36 & 0.87 & 0.36 & 0.92 & 0.61 & 0.9 & 0.61 & 0.51 & 0.5 \\
$Q_1$ & 0.98 & 0.73 & 0.97 & 0.74 & 0.96 & 0.84 & 0.96 & 0.87 & 0.84 & 0.96\\
$Q_2$ & 1 & 0.88 & 0.99 & 0.88 & 0.99 & 0.92 & 0.98 & 0.92 & 0.9 & 1\\
$Q_3$ & 1.02 & 0.91 & 1.01 & 0.91 & 1 & 0.95 & 1 & 0.94 & 0.96 & 1.04\\
MAX & 1.13 & 0.95 & 1.15 & 0.96 & 1.01 & 0.97 & 1.12 & 0.98 & 1.04 & 1.48\\

\end{tabular}\\
\vspace{.5em}
\centering
\rotatebox[origin=c]{90}{AMBA/GenBuf\hspace{1em}}
\rotatebox[origin=c]{90}{unrealizable\hspace{1em}}
\hspace{1em}
\begin{tabular}{l | r r r r | r r r r | r r}
~~ & \multicolumn{4}{|c|}{GR(1) algorithm} & \multicolumn{4}{|c|}{Rabin(1) algorithm} &
\multicolumn{2}{|c}{Rabin(1) / GR(1)}\\
\hline
Quartile & efp & eun & fpr & all & efp & eun & fpr & all & orig & all\\
\hline
MIN & 0.85 & 0.001 & 0.93 & 0.001 & 0.71 & 0.001 & 0.89 & 0.001 & 0.68 & 0.69\\
$Q_1$ & 0.99 & 0.1 & 0.99 & 0.1 & 0.96 & 0.09 & 0.98 & 0.09 & 0.87 & 0.93\\
$Q_2$ & 1 & 0.54 & 1 & 0.52 & 1 & 0.62 & 0.99 & 0.57 & 0.93 & 0.97\\
$Q_3$ & 1  & 0.97 & 1.02 & 0.97 & 1 & 0.98 & 1 & 0.98 & 1 & 1.01\\
MAX & 1.33 & 1.07  & 1.06 & 1.07 & 1.3 & 1.01 & 1.03 & 1.01 & 1.85 & 1.86\\

\end{tabular}
\caption{\small Ratios of the heuristics to the original GR(1)
and Rabin(1) running times for unrealizable specifications.}
\label{tbl:gr1_unreal}
\end{table}

\paragraph{RQ1: Effectiveness of heuristics}
The heuristics of early fixed-point detection reduces running times by
at least 5\% on 25\% of the realizable specifications
(Tbl.~\ref{tbl:gr1_real}, \emph{efp}), but seems even less effective on
unrealizable specifications (Tbl.~\ref{tbl:gr1_unreal}, \emph{efp}). As
expected, the early detection of unrealizability has no notable effect
on realizable specifications (Tbl.~\ref{tbl:gr1_real}, \emph{eun}), but
on unrealizable specifications reduces running times of 50\% of the
specifications by at least 12\%/46\% for GR(1) and more than 8\%/38\%
for Rabin(1) (Tbl.~\ref{tbl:gr1_real}, \emph{eun}). The heuristics of
fixed-point recycling appears ineffective for unrealizable
specifications (Tbl.~\ref{tbl:gr1_unreal}), but reduces running times of
25\% of the realizable specifications by at least 7\%/17\% for GR(1) and
at least 6\%/14\% for Rabin(1) (Tbl.~\ref{tbl:gr1_real}, \emph{fpr}).
As good news, the combination of all heuristics usually improves over
each heuristics separately (column \emph{all}). Another interesting
observation is that the Rabin(1) algorithm determines realizability faster
than the GR(1) algorithm for almost all specifications.

\paragraph{RQ2: Difference between specification sets}
For realizable specifications, we see that the suggested heuristics
perform better on the AMBA and GenBuf set than on \wsSpecs, i.e., all
heuristics (columns \emph{all}) decreases running times on 50\% of the
AMBA and GenBuf specifications by at least 10\% and for \wsSpecs
specifications by at least 5\%. A more significant difference between
the specification sets is revealed by Tbl.~\ref{tbl:gr1_unreal} of
unrealizable specifications. Here the speedup for 50\% of the
specifications, mainly obtained by \emph{eun}, is at least around 10\%
for \wsSpecs but at least around 50\% for AMBA and GenBuf. We believe
that this difference is due to the systematic and synthetic reasons for
unrealizability added by Cimatti et al.~\cite{CimattiRST08}.

\subsubsection{Results for Unrealizable Core Calculation}

We present the ratios of running times for heuristics from
Sect.~\ref{sec:heuristics:core} for unrealizable specifications from the
sets \wsSpecs and AMBA and GenBuf. The different heuristics are
abbreviated as follows: \emph{sets} is the contained sets in the core
calculation from Sect.~\ref{sec:ddmin:containedSets}, \emph{opt} uses
the optimized GR(1) and Rabin(1) algorithms from
Sect.~\ref{sec:heuristics:algorithms}, and \emph{inc} is the incremental
algorithm for similar candidates from Sect.~\ref{sec:ddmin:incremental}.
Here, by \emph{all} we refer to the combination of \emph{sets} and
\emph{opt} but not \emph{inc}, because only the first two seem to improve
running times.
All the results are rounded to two decimals (or more if otherwise 0).
Tbl.~\ref{tbl:ddmin} shows the ratios of running times for 17
unrealizable \wsSpecs specifications (top) and for 30 unrealizable AMBA
and GenBuf specifications (bottom). All tables show first ratios of
running times for \ddmin with the GR(1) algorithm, then ratios for
\ddmin with the Rabin(1) algorithm, and finally a comparison between the
Rabin(1) and GR(1) algorithms.

\begin{table}[t]\small
\centering
\rotatebox[origin=c]{90}{\wsSpecs\hspace{1em}}
\rotatebox[origin=c]{90}{unrealizable\hspace{1em}}
\hspace{1em}
\begin{tabular}{l | r r r r | r r r r | r r}
~~ & \multicolumn{4}{|c|}{DDmin with GR(1)} & \multicolumn{4}{|c|}{DDmin
with Rabin(1)} & \multicolumn{2}{|c}{Rabin(1) / GR(1)}\\
\hline
Quartile & sets & opt & inc & all & sets & opt & inc & all & orig & all\\
\hline
MIN & 0.47 & 0.66 & 0.79 & 0.3 & 0.44 & 0.75 & 0.73 & 0.35 & 0.85 &  0.89 \\
$Q_1$ & 0.56 & 0.94 & 1.19 & 0.5 & 0.59 & 0.92 & 1.32 & 0.51 & 1.03 & 1.04\\
$Q_2$ & 0.6 & 0.96 & 1.32 & 0.56 & 0.65 & 0.95 & 1.49 & 0.55 & 1.05 & 1.09 \\
$Q_3$ & 0.73 & 0.97 & 1.62 & 0.6 & 0.74 & 0.98 & 1.65 & 0.65 & 1.19 & 1.28\\
MAX & 0.75 & 0.98 & 2 & 0.71 & 0.78 & 1.03 & 2.11 & 0.78 & 1.38 & 1.85 \\

\end{tabular}\\
\vspace{.5em}
\centering
\rotatebox[origin=c]{90}{AMBA/GenBuf\hspace{1em}}
\rotatebox[origin=c]{90}{unrealizable\hspace{1em}}
\hspace{1em}
\begin{tabular}{l | r r r r | r r r r | r r}
~~ & \multicolumn{4}{|c|}{DDmin with GR(1)} & \multicolumn{4}{|c|}{DDmin
with Rabin(1)} & \multicolumn{2}{|c}{Rabin(1) / GR(1)}\\
\hline
Quartile & sets & opt & inc & all & sets & opt & inc & all & orig & all\\
\hline
MIN & 0.46 & 0.05 & 0.91 & 0.02 & 0.46 &  0.04 & 0.69 & 0.02 & 0.66 & 0.81\\
$Q_1$ & 0.61 & 0.71 & 1.08 & 0.45 & 0.61 & 0.72 & 1.09 & 0.45 & 0.93 & 0.93\\
$Q_2$ & 0.69 & 0.9 & 1.35 & 0.57 & 0.7 & 0.94 & 1.28 & 0.56 & 1.02 & 1.01\\
$Q_3$ & 0.91 & 0.97 & 1.46 & 0.66  & 0.83 & 0.97 & 1.64 & 0.65 & 1.13 & 1.18\\
MAX & 1.2 & 1.12 & 2.23 & 1.09 & 3.08 &  1.06 & 2.38 & 0.91 & 1.69 & 1.41\\

\end{tabular}
\caption{\small Ratios of the heuristics to the original \ddmin running
times for unrealizable specifications.}
\label{tbl:ddmin}
\end{table}

\paragraph{RQ1: Effectiveness of heuristics}
The heuristics of contained sets appears very effective on all
specifications and reduces running times of 50\% of the specifications
by at least 40\%/31\% for \ddmin with GR(1) and at least 35\%/30\% for
\ddmin with Rabin(1) (Tbl.~\ref{tbl:ddmin}, \emph{sets}). Using the GR(1)
and Rabin(1) algorithms with all heuristics again improves running times
for 50\% of the specifications by at least 4\% (columns \emph{opt}).
Contrary to our expectation the reuse of previous BDDs for incremental
game solving slows down running times on almost all specifications
(columns \emph{inc}) with a maximum factor of 2.38x. We believe that
this increase in running times is due to increased BDD
variable reordering times.
We use the automatic reorder of CUDD in all our tests, and the overall reordering time is
directly affected by keeping many BDDs of previous runs. 
As good news again, the combination of the heuristics \emph{sets} and \emph{opt} usually improves running times even further
and roughly obtains a speedup of at least 2x for 50\% of the
specifications (column \emph{all}).

\paragraph{RQ2: Difference between specification sets}
The combination of all heuristics similarly improves running times for
the \wsSpecs and the AMBA and GenBuf specifications (columns
\emph{all}). The heuristics \emph{sets} consistently performs a
few percent better on \wsSpecs than on AMBA and GenBuf. The heuristics \emph{opt} performs
better on the first quartile of AMBA and GenBuf specifications. This is
consistent with the observed behavior in Tbl.~\ref{tbl:gr1_unreal}. 


\subsection{Validation of Heuristics' Correctness}
\label{sec:validation}
Our implementation of the different heuristics might have bugs, so to
ensure correctness of the code we performed the following validation. 
We have computed the complete set of winning states using the
original algorithm and compared the result to the winning states
computed by the modified algorithms employing each of the three heuristics
separately. As expected, only for unrealizable specifications
the heuristics for detecting unrealizability early computed less
winning states.

To further ensure that the game memory allows
for strategy construction (memory is different for fixed-point
recycling), we have synthesized strategies from the game memory produced when using our
heuristics. We have verified the correctness of the strategies by
LTL model checking against the LTL specifications for strict
realizability of the original GR(1) and Rabin(1) specifications.

For the \ddmin heuristics, we have compared the resulting core of each
heuristics to the original one. Since the heuristics are not on the \ddmin
itself but on the \texttt{check}, the core was never different.
Furthermore, we executed \ddmin again on the core, to
validate the local minimum.

Validation was successful on all 118  specifications used in this
paper.

\subsection{Threats to Validity}
\label{sec:threats}
We discuss threats to the validity of our results.

\noindent\textbf{Internal.}
The implementation of the different heuristics might have bugs, so to ensure
correctness of the code we performed validations as described in Sect.~\ref{sec:validation}.

Another threat is the variation of the running times of the same test. Different
runs of the same algorithm may result in slightly different running times, so
the ratios we showed in Sect.~\ref{sect:evaluation} might not be accurate if we
run each test only once. We mitigate it by performing 50 runs of each algorithm and reporting medians
as described in Sect.~\ref{sec:procedure}.

\noindent\textbf{External.} The results of the different heuristics
might not be generalizable due to the limited number of specifications
used in our evaluation.
We divided our evaluation into two sets: (1) \wsSpecs, which are
realistic specifications created by students for different robotic
systems, and (2) the AMBA and GenBuf specifications, which were created
by researchers and systematically scaled to larger sizes.
The total number of the specifications might be insufficient. The set
\wsSpecs consists of 78 specifications (17 unrealizable).
The set AMBA and GenBuf consists of 40 specifications (30 unrealizable). 

We share some observations on the sets of specifications that might have an
influence of generalizability of the results.
First, the AMBA and GenBuf specifications used in literature were generated
systematically for growing parameters (number of AMBA arbiters and GenBuf
requests). Thus the 40 AMBA and GenBuf specifications essentially describe only
two systems. Furthermore, the reasons for unrealizability of AMBA and GenBuf
were systematically introduced~\cite{CimattiRST08} and consist of a single
change each.
Second, the running times of checking realizability of the \wsSpecs
specifications are rather low and range from 1.5ms to 1300ms, with median around
30ms. In this set the specifications are biased based on the numbers of
revisions committed by students: the Humanoid has 21 specifications (8
unrealizable), the Gyro has 11 specifications (2 unrealizable), and the
SelfParkingCar has only 4 specifications in total.
Furthermore, none of the specifications were written by engineers, so we cannot
evaluate how our results may generalize to large scale real-world
specifications.

\section{Related Work}

Könighofer et al.~\cite{KonighoferHB13} presented diagnoses for
unrealizable GR(1) specifications. They also implemented the heuristics
for \ddmin mentioned in Sect.~\ref{sec:ddmin:containedSets}. They suggest
further heuristics that approximate the set of system winning states.
These heuristics are different from the ones we presented as they are 
riskier: in case they fail the computation reverts to the original GR(1)
algorithm. An analysis of the speed-up obtained from their heuristics
for \ddmin alone was not reported.

Others have focused on strategy construction for GR(1). Strategies are
constructed from the memory stored in the \texttt{X}, \texttt{Y}, and
\texttt{Z} arrays in Alg.~\ref{alg:gr1} and Alg.~\ref{alg:rabin}.
Schlaipfer et al.~\cite{SchlaipferHB11} suggest synthesis of separate
strategies for each justice guarantee to avoid a blow-up of the BDD
representation. Bloem et al.~\cite{BJP+12} discuss different
minimization of synthesized strategies that do not necessarily minimize
their BDDs.
We consider space and time related heuristics for strategy construction
an interesting next step.

It is well-known that the order of BDD variables heavily influences the
performance of BDD-based
algorithms~\cite{JacobsBBEHKPRRS15,YangBOBCJRS98}. The GR(1)
implementation of Slugs~\cite{EhlersR16} uses the default dynamic
variable reordering of CUDD~\cite{CUDD} (as we do). Slugs turns off
reordering during strategy construction. Filippidis et
al.~\cite{FilippidisMH16} reported better performance with reordering
during strategy construction. We are not aware of any GR(1) specific
heuristics for (dynamic) BDD variable ordering.

As a very different and complementary approach to ours, one can consider
rewriting the GR(1) specification to speed up realizability checking and
synthesis. Filippidis et al.~\cite{FilippidisMH16} report on obtaining a
speedup of factor 100 for synthesizing AMBA by manually changing the
AMBA specification of~\cite{BJP+12} to use less variables and weaker
assumptions. We have not focused on these very specific optimizations of
single specifications.
Our work presents and evaluates specification agnostic heuristics.

Finally, a number of heuristics for BDD-based safety game solvers have
been reported as outcome of the SYNTCOMP reactive synthesis
competitions~\cite{JacobsBBEHKPRRS15,JacobsBBK0KKLNP16,JacobsBBKPRRSST16}.
Most of these optimizations are on the level of predecessor
computations (operators $\circlediamond$ in Alg.~\ref{alg:gr1} and
$\circlebox$ in Alg.~\ref{alg:rabin}), while the heuristics we
implemented are on the level of fixed-points and repeated computations.
It seems possible to combine these heuristics. Notably, an approach for
predicate abstraction for predecessor computation has already been
implemented for GR(1) synthesis~\cite{RyzhykW16,WalkerR14}.

\section{Conclusion}
\label{sec:conclusion}

We presented a list of heuristics to potentially reduce running times
for GR(1) synthesis and related algorithms. The list includes early
detection of fixed-points and unrealizability, fixed-point recycling,
and heuristics for unrealizable core computations.
We implemented and evaluated the heuristics and their combination on two
sets of benchmarks, first \wsSpecs, a set of 78 specifications created
by 3rd year undergraduate computer science students in a project class
of one semester, and second on the two systems AMBA and GenBuf available
and well-studied in GR(1) literature.

Our evaluation shows that most heuristics have a positive effect on
running times for checking realizability of a specification and for
unrealizable core calculation. Most importantly, their combination
outperforms the individual heuristics and even in the worst-case has no
or a very low overhead. In addition, the heuristics similarly improve
running times for both sets of specifications whereas the synthetic
reasons for unrealizability in AMBA and GenBuf lead to faster
computations.

The work is part of a larger project on bridging the gap between the
theory and algorithms of reactive synthesis on the one hand and software
engineering practice on the other.  As part of this project we are
building engineer-friendly tools for reactive synthesis, see,
e.g.,~\cite{MPR16,MaozR15,MaozR15synt,MR16}.

\paragraph{Acknowledgments}
This project has received funding from the
European Research Council (ERC) under the European Union's Horizon 2020
research and innovation programme (grant agreement No 638049, SYNTECH).

\bibliographystyle{eptcs}
\bibliography{doc}

\begin{thebibliography}{10}
\providecommand{\bibitemdeclare}[2]{}
\providecommand{\surnamestart}{}
\providecommand{\surnameend}{}
\providecommand{\urlprefix}{Available at }
\providecommand{\url}[1]{\texttt{#1}}
\providecommand{\href}[2]{\texttt{#2}}
\providecommand{\urlalt}[2]{\href{#1}{#2}}
\providecommand{\doi}[1]{doi:\urlalt{http://dx.doi.org/#1}{#1}}
\providecommand{\bibinfo}[2]{#2}

\bibitemdeclare{article}{BJP+12}
\bibitem{BJP+12}
\bibinfo{author}{Roderick \surnamestart Bloem\surnameend},
  \bibinfo{author}{Barbara \surnamestart Jobstmann\surnameend},
  \bibinfo{author}{Nir \surnamestart Piterman\surnameend},
  \bibinfo{author}{Amir \surnamestart Pnueli\surnameend} \&
  \bibinfo{author}{Yaniv \surnamestart Sa'ar\surnameend}
  (\bibinfo{year}{2012}): \emph{\bibinfo{title}{{Synthesis of Reactive(1)
  Designs}}}.
\newblock {\sl \bibinfo{journal}{J. Comput. Syst. Sci.}}
  \bibinfo{volume}{78}(\bibinfo{number}{3}), pp. \bibinfo{pages}{911--938},
  \doi{10.1016/j.jcss.2011.08.007}.

\bibitemdeclare{article}{BrowneCJLM97}
\bibitem{BrowneCJLM97}
\bibinfo{author}{Anca \surnamestart Browne\surnameend},
  \bibinfo{author}{Edmund~M. \surnamestart Clarke\surnameend},
  \bibinfo{author}{Somesh \surnamestart Jha\surnameend},
  \bibinfo{author}{David~E. \surnamestart Long\surnameend} \&
  \bibinfo{author}{Wilfredo~R. \surnamestart Marrero\surnameend}
  (\bibinfo{year}{1997}): \emph{\bibinfo{title}{An Improved Algorithm for the
  Evaluation of Fixpoint Expressions}}.
\newblock {\sl \bibinfo{journal}{Theor. Comput. Sci.}}
  \bibinfo{volume}{178}(\bibinfo{number}{1-2}), pp. \bibinfo{pages}{237--255},
  \doi{10.1016/S0304-3975(96)00228-9}.

\bibitemdeclare{proceedings}{DBLP:journals/corr/CernyKM16}
\bibitem{DBLP:journals/corr/CernyKM16}
\bibinfo{editor}{Pavol \surnamestart Cern{\'{y}}\surnameend},
  \bibinfo{editor}{Viktor \surnamestart Kuncak\surnameend} \&
  \bibinfo{editor}{Parthasarathy \surnamestart Madhusudan\surnameend}, editors
  (\bibinfo{year}{2016}): \emph{\bibinfo{title}{Proceedings Fourth Workshop on
  Synthesis, {SYNT} 2015, San Francisco, CA, USA, 18th July 2015}}. {\sl
  \bibinfo{series}{{EPTCS}}} \bibinfo{volume}{202}, \doi{10.4204/EPTCS.202}.

\bibitemdeclare{inproceedings}{CimattiRST08}
\bibitem{CimattiRST08}
\bibinfo{author}{Alessandro \surnamestart Cimatti\surnameend},
  \bibinfo{author}{Marco \surnamestart Roveri\surnameend},
  \bibinfo{author}{Viktor \surnamestart Schuppan\surnameend} \&
  \bibinfo{author}{Andrei \surnamestart Tchaltsev\surnameend}
  (\bibinfo{year}{2008}): \emph{\bibinfo{title}{Diagnostic Information for
  Realizability}}.
\newblock In: {\sl \bibinfo{booktitle}{VMCAI}}, {\sl \bibinfo{series}{LNCS}}
  \bibinfo{volume}{4905}, \bibinfo{publisher}{Springer}, pp.
  \bibinfo{pages}{52--67}, \doi{10.1007/978-3-540-78163-9_9}.

\bibitemdeclare{article}{DIppolitoBPU13}
\bibitem{DIppolitoBPU13}
\bibinfo{author}{Nicol{\'{a}}s \surnamestart D'Ippolito\surnameend},
  \bibinfo{author}{V{\'{\i}}ctor~A. \surnamestart Braberman\surnameend},
  \bibinfo{author}{Nir \surnamestart Piterman\surnameend} \&
  \bibinfo{author}{Sebasti{\'{a}}n \surnamestart Uchitel\surnameend}
  (\bibinfo{year}{2013}): \emph{\bibinfo{title}{Synthesizing nonanomalous
  event-based controllers for liveness goals}}.
\newblock {\sl \bibinfo{journal}{{ACM} Trans. Softw. Eng. Methodol.}}
  \bibinfo{volume}{22}(\bibinfo{number}{1}), p.~\bibinfo{pages}{9},
  \doi{10.1145/2430536.2430543}.

\bibitemdeclare{inproceedings}{DAC99}
\bibitem{DAC99}
\bibinfo{author}{Matthew~B. \surnamestart Dwyer\surnameend},
  \bibinfo{author}{George~S. \surnamestart Avrunin\surnameend} \&
  \bibinfo{author}{James~C. \surnamestart Corbett\surnameend}
  (\bibinfo{year}{1999}): \emph{\bibinfo{title}{Patterns in Property
  Specifications for Finite-State Verification}}.
\newblock In: {\sl \bibinfo{booktitle}{ICSE}}, \bibinfo{publisher}{{ACM}}, pp.
  \bibinfo{pages}{411--420}, \doi{10.1145/302405.302672}.

\bibitemdeclare{inproceedings}{Ehlers11}
\bibitem{Ehlers11}
\bibinfo{author}{R{\"{u}}diger \surnamestart Ehlers\surnameend}
  (\bibinfo{year}{2011}): \emph{\bibinfo{title}{Generalized {R}abin(1)
  Synthesis with Applications to Robust System Synthesis}}.
\newblock In: {\sl \bibinfo{booktitle}{{NASA} Formal Methods}}, {\sl
  \bibinfo{series}{LNCS}} \bibinfo{volume}{6617},
  \bibinfo{publisher}{Springer}, pp. \bibinfo{pages}{101--115},
  \doi{10.1007/978-3-642-20398-5\_9}.

\bibitemdeclare{inproceedings}{EhlersR16}
\bibitem{EhlersR16}
\bibinfo{author}{R{\"{u}}diger \surnamestart Ehlers\surnameend} \&
  \bibinfo{author}{Vasumathi \surnamestart Raman\surnameend}
  (\bibinfo{year}{2016}): \emph{\bibinfo{title}{Slugs: Extensible {GR(1)}
  Synthesis}}.
\newblock In \bibinfo{editor}{Swarat \surnamestart Chaudhuri\surnameend} \&
  \bibinfo{editor}{Azadeh \surnamestart Farzan\surnameend}, editors: {\sl
  \bibinfo{booktitle}{Computer Aided Verification - 28th International
  Conference, {CAV} 2016, Toronto, ON, Canada, July 17-23, 2016, Proceedings,
  Part {II}}}, {\sl \bibinfo{series}{Lecture Notes in Computer Science}}
  \bibinfo{volume}{9780}, \bibinfo{publisher}{Springer}, pp.
  \bibinfo{pages}{333--339}, \doi{10.1007/978-3-319-41540-6_18}.

\bibitemdeclare{inproceedings}{FilippidisMH16}
\bibitem{FilippidisMH16}
\bibinfo{author}{Ioannis \surnamestart Filippidis\surnameend},
  \bibinfo{author}{Richard~M. \surnamestart Murray\surnameend} \&
  \bibinfo{author}{Gerard~J. \surnamestart Holzmann\surnameend}
  (\bibinfo{year}{2015}): \emph{\bibinfo{title}{A multi-paradigm language for
  reactive synthesis}}.
\newblock In \bibinfo{editor}{Cern{\'{y}}} et~al.
  \cite{DBLP:journals/corr/CernyKM16}, pp. \bibinfo{pages}{73--97},
  \doi{10.4204/EPTCS.202.6}.

\bibitemdeclare{proceedings}{2001automata}
\bibitem{2001automata}
\bibinfo{editor}{Erich \surnamestart Gr{\"{a}}del\surnameend},
  \bibinfo{editor}{Wolfgang \surnamestart Thomas\surnameend} \&
  \bibinfo{editor}{Thomas \surnamestart Wilke\surnameend}, editors
  (\bibinfo{year}{2002}): \emph{\bibinfo{title}{Automata, Logics, and Infinite
  Games: {A} Guide to Current Research [outcome of a Dagstuhl seminar, February
  2001]}}. {\sl \bibinfo{series}{Lecture Notes in Computer Science}}
  \bibinfo{volume}{2500}, \bibinfo{publisher}{Springer},
  \doi{10.1007/3-540-36387-4}.

\bibitemdeclare{article}{JacobsBBEHKPRRS15}
\bibitem{JacobsBBEHKPRRS15}
\bibinfo{author}{Swen \surnamestart Jacobs\surnameend},
  \bibinfo{author}{Roderick \surnamestart Bloem\surnameend},
  \bibinfo{author}{Romain \surnamestart Brenguier\surnameend},
  \bibinfo{author}{R{\"{u}}diger \surnamestart Ehlers\surnameend},
  \bibinfo{author}{Timotheus \surnamestart Hell\surnameend},
  \bibinfo{author}{Robert \surnamestart K{\"{o}}nighofer\surnameend},
  \bibinfo{author}{Guillermo~A. \surnamestart P{\'{e}}rez\surnameend},
  \bibinfo{author}{Jean{-}Fran{\c{c}}ois \surnamestart Raskin\surnameend},
  \bibinfo{author}{Leonid \surnamestart Ryzhyk\surnameend},
  \bibinfo{author}{Ocan \surnamestart Sankur\surnameend},
  \bibinfo{author}{Martina \surnamestart Seidl\surnameend},
  \bibinfo{author}{Leander \surnamestart Tentrup\surnameend} \&
  \bibinfo{author}{Adam \surnamestart Walker\surnameend}
  (\bibinfo{year}{2017}): \emph{\bibinfo{title}{The first reactive synthesis
  competition {(SYNTCOMP} 2014)}}.
\newblock {\sl \bibinfo{journal}{{STTT}}}
  \bibinfo{volume}{19}(\bibinfo{number}{3}), pp. \bibinfo{pages}{367--390},
  \doi{10.1007/s10009-016-0416-3}.

\bibitemdeclare{inproceedings}{JacobsBBK0KKLNP16}
\bibitem{JacobsBBK0KKLNP16}
\bibinfo{author}{Swen \surnamestart Jacobs\surnameend},
  \bibinfo{author}{Roderick \surnamestart Bloem\surnameend},
  \bibinfo{author}{Romain \surnamestart Brenguier\surnameend},
  \bibinfo{author}{Ayrat \surnamestart Khalimov\surnameend},
  \bibinfo{author}{Felix \surnamestart Klein\surnameend},
  \bibinfo{author}{Robert \surnamestart K{\"{o}}nighofer\surnameend},
  \bibinfo{author}{Jens \surnamestart Kreber\surnameend},
  \bibinfo{author}{Alexander \surnamestart Legg\surnameend},
  \bibinfo{author}{Nina \surnamestart Narodytska\surnameend},
  \bibinfo{author}{Guillermo~A. \surnamestart P{\'{e}}rez\surnameend},
  \bibinfo{author}{Jean{-}Fran{\c{c}}ois \surnamestart Raskin\surnameend},
  \bibinfo{author}{Leonid \surnamestart Ryzhyk\surnameend},
  \bibinfo{author}{Ocan \surnamestart Sankur\surnameend},
  \bibinfo{author}{Martina \surnamestart Seidl\surnameend},
  \bibinfo{author}{Leander \surnamestart Tentrup\surnameend} \&
  \bibinfo{author}{Adam \surnamestart Walker\surnameend}
  (\bibinfo{year}{2016}): \emph{\bibinfo{title}{The 3rd Reactive Synthesis
  Competition {(SYNTCOMP} 2016): Benchmarks, Participants {\&} Results}}.
\newblock In \bibinfo{editor}{Piskac} \& \bibinfo{editor}{Dimitrova}
  \cite{DBLP:journals/corr/PiskacD16}, pp. \bibinfo{pages}{149--177},
  \doi{10.4204/EPTCS.229.12}.

\bibitemdeclare{inproceedings}{JacobsBBKPRRSST16}
\bibitem{JacobsBBKPRRSST16}
\bibinfo{author}{Swen \surnamestart Jacobs\surnameend},
  \bibinfo{author}{Roderick \surnamestart Bloem\surnameend},
  \bibinfo{author}{Romain \surnamestart Brenguier\surnameend},
  \bibinfo{author}{Robert \surnamestart K{\"{o}}nighofer\surnameend},
  \bibinfo{author}{Guillermo~A. \surnamestart P{\'{e}}rez\surnameend},
  \bibinfo{author}{Jean{-}Fran{\c{c}}ois \surnamestart Raskin\surnameend},
  \bibinfo{author}{Leonid \surnamestart Ryzhyk\surnameend},
  \bibinfo{author}{Ocan \surnamestart Sankur\surnameend},
  \bibinfo{author}{Martina \surnamestart Seidl\surnameend},
  \bibinfo{author}{Leander \surnamestart Tentrup\surnameend} \&
  \bibinfo{author}{Adam \surnamestart Walker\surnameend}
  (\bibinfo{year}{2015}): \emph{\bibinfo{title}{The Second Reactive Synthesis
  Competition {(SYNTCOMP} 2015)}}.
\newblock In \bibinfo{editor}{Cern{\'{y}}} et~al.
  \cite{DBLP:journals/corr/CernyKM16}, pp. \bibinfo{pages}{27--57},
  \doi{10.4204/EPTCS.202.4}.

\bibitemdeclare{article}{KonighoferHB13}
\bibitem{KonighoferHB13}
\bibinfo{author}{Robert \surnamestart K{\"{o}}nighofer\surnameend},
  \bibinfo{author}{Georg \surnamestart Hofferek\surnameend} \&
  \bibinfo{author}{Roderick \surnamestart Bloem\surnameend}
  (\bibinfo{year}{2013}): \emph{\bibinfo{title}{Debugging formal
  specifications: a practical approach using model-based diagnosis and
  counterstrategies}}.
\newblock {\sl \bibinfo{journal}{{STTT}}}
  \bibinfo{volume}{15}(\bibinfo{number}{5-6}), pp. \bibinfo{pages}{563--583},
  \doi{10.1007/s10009-011-0221-y}.

\bibitemdeclare{article}{Kozen83}
\bibitem{Kozen83}
\bibinfo{author}{Dexter \surnamestart Kozen\surnameend} (\bibinfo{year}{1983}):
  \emph{\bibinfo{title}{Results on the Propositional mu-Calculus}}.
\newblock {\sl \bibinfo{journal}{Theor. Comput. Sci.}} \bibinfo{volume}{27},
  pp. \bibinfo{pages}{333--354}, \doi{10.1016/0304-3975(82)90125-6}.

\bibitemdeclare{article}{Kress-GazitFP09}
\bibitem{Kress-GazitFP09}
\bibinfo{author}{Hadas \surnamestart Kress{-}Gazit\surnameend},
  \bibinfo{author}{Georgios~E. \surnamestart Fainekos\surnameend} \&
  \bibinfo{author}{George~J. \surnamestart Pappas\surnameend}
  (\bibinfo{year}{2009}): \emph{\bibinfo{title}{Temporal-Logic-Based Reactive
  Mission and Motion Planning}}.
\newblock {\sl \bibinfo{journal}{{IEEE} Trans. Robotics}}
  \bibinfo{volume}{25}(\bibinfo{number}{6}), pp. \bibinfo{pages}{1370--1381},
  \doi{10.1109/TRO.2009.2030225}.

\bibitemdeclare{proceedings}{DBLP:journals/taosd/2013-10}
\bibitem{DBLP:journals/taosd/2013-10}
\bibinfo{editor}{Gary~T. \surnamestart Leavens\surnameend},
  \bibinfo{editor}{Shigeru \surnamestart Chiba\surnameend} \&
  \bibinfo{editor}{{\'{E}}ric \surnamestart Tanter\surnameend}, editors
  (\bibinfo{year}{2013}): \emph{\bibinfo{title}{Transactions on Aspect-Oriented
  Software Development {X}}}. {\sl \bibinfo{series}{Lecture Notes in Computer
  Science}} \bibinfo{volume}{7800}, \bibinfo{publisher}{Springer},
  \doi{10.1007/978-3-642-36964-3}.

\bibitemdeclare{inproceedings}{MPR16}
\bibitem{MPR16}
\bibinfo{author}{Shahar \surnamestart Maoz\surnameend},
  \bibinfo{author}{Or~\surnamestart Pistiner\surnameend} \&
  \bibinfo{author}{Jan~Oliver \surnamestart Ringert\surnameend}
  (\bibinfo{year}{2016}): \emph{\bibinfo{title}{Symbolic {BDD} and {ADD}
  Algorithms for Energy Games}}.
\newblock In \bibinfo{editor}{Piskac} \& \bibinfo{editor}{Dimitrova}
  \cite{DBLP:journals/corr/PiskacD16}, pp. \bibinfo{pages}{35--54},
  \doi{10.4204/EPTCS.229.5}.

\bibitemdeclare{inproceedings}{MaozR15}
\bibitem{MaozR15}
\bibinfo{author}{Shahar \surnamestart Maoz\surnameend} \&
  \bibinfo{author}{Jan~Oliver \surnamestart Ringert\surnameend}
  (\bibinfo{year}{2015}): \emph{\bibinfo{title}{{GR(1)} synthesis for {LTL}
  specification patterns}}.
\newblock In \bibinfo{editor}{Elisabetta~Di \surnamestart Nitto\surnameend},
  \bibinfo{editor}{Mark \surnamestart Harman\surnameend} \&
  \bibinfo{editor}{Patrick \surnamestart Heymans\surnameend}, editors: {\sl
  \bibinfo{booktitle}{Proceedings of the 2015 10th Joint Meeting on Foundations
  of Software Engineering, {ESEC/FSE} 2015, Bergamo, Italy, August 30 -
  September 4, 2015}}, \bibinfo{publisher}{{ACM}}, pp.
  \bibinfo{pages}{96--106}, \doi{10.1145/2786805.2786824}.

\bibitemdeclare{inproceedings}{MaozR15synt}
\bibitem{MaozR15synt}
\bibinfo{author}{Shahar \surnamestart Maoz\surnameend} \&
  \bibinfo{author}{Jan~Oliver \surnamestart Ringert\surnameend}
  (\bibinfo{year}{2015}): \emph{\bibinfo{title}{{Synthesizing a Lego Forklift
  Controller in GR(1): A Case Study}}}.
\newblock In: {\sl \bibinfo{booktitle}{Proc. 4th Workshop on Synthesis, {SYNT}
  2015 colocated with CAV 2015}}, {\sl \bibinfo{series}{{EPTCS}}}
  \bibinfo{volume}{202}, pp. \bibinfo{pages}{58--72},
  \doi{10.4204/EPTCS.202.5}.

\bibitemdeclare{inproceedings}{MR16}
\bibitem{MR16}
\bibinfo{author}{Shahar \surnamestart Maoz\surnameend} \&
  \bibinfo{author}{Jan~Oliver \surnamestart Ringert\surnameend}
  (\bibinfo{year}{2016}): \emph{\bibinfo{title}{On well-separation of {GR(1)}
  specifications}}.
\newblock In \bibinfo{editor}{Thomas \surnamestart Zimmermann\surnameend},
  \bibinfo{editor}{Jane \surnamestart Cleland{-}Huang\surnameend} \&
  \bibinfo{editor}{Zhendong \surnamestart Su\surnameend}, editors: {\sl
  \bibinfo{booktitle}{Proceedings of the 24th {ACM} {SIGSOFT} International
  Symposium on Foundations of Software Engineering, {FSE} 2016, Seattle, WA,
  USA, November 13-18, 2016}}, \bibinfo{publisher}{{ACM}}, pp.
  \bibinfo{pages}{362--372}, \doi{10.1145/2950290.2950300}.

\bibitemdeclare{inproceedings}{MaozS11}
\bibitem{MaozS11}
\bibinfo{author}{Shahar \surnamestart Maoz\surnameend} \&
  \bibinfo{author}{Yaniv \surnamestart Sa'ar\surnameend}
  (\bibinfo{year}{2011}): \emph{\bibinfo{title}{{AspectLTL}: an aspect language
  for {LTL} specifications}}.
\newblock In \bibinfo{editor}{Paulo \surnamestart Borba\surnameend} \&
  \bibinfo{editor}{Shigeru \surnamestart Chiba\surnameend}, editors: {\sl
  \bibinfo{booktitle}{AOSD}}, \bibinfo{publisher}{{ACM}}, pp.
  \bibinfo{pages}{19--30}, \doi{10.1145/1960275.1960280}.

\bibitemdeclare{inproceedings}{MaozS12}
\bibitem{MaozS12}
\bibinfo{author}{Shahar \surnamestart Maoz\surnameend} \&
  \bibinfo{author}{Yaniv \surnamestart Sa'ar\surnameend}
  (\bibinfo{year}{2012}): \emph{\bibinfo{title}{Assume-Guarantee Scenarios:
  Semantics and Synthesis}}.
\newblock In: {\sl \bibinfo{booktitle}{MODELS}}, {\sl \bibinfo{series}{LNCS}}
  \bibinfo{volume}{7590}, \bibinfo{publisher}{Springer}, pp.
  \bibinfo{pages}{335--351}, \doi{10.1007/978-3-642-33666-9\_22}.

\bibitemdeclare{article}{MaozS13AOSD}
\bibitem{MaozS13AOSD}
\bibinfo{author}{Shahar \surnamestart Maoz\surnameend} \&
  \bibinfo{author}{Yaniv \surnamestart Sa'ar\surnameend}
  (\bibinfo{year}{2013}): \emph{\bibinfo{title}{Two-Way Traceability and
  Conflict Debugging for AspectLTL Programs}}.
\newblock In {\sl \bibinfo{journal}{T. Aspect-Oriented Software Development}}
  \cite{DBLP:journals/taosd/2013-10}, pp. \bibinfo{pages}{39--72},
  \doi{10.1007/978-3-642-36964-3\_2}.

\bibitemdeclare{proceedings}{DBLP:journals/corr/PiskacD16}
\bibitem{DBLP:journals/corr/PiskacD16}
\bibinfo{editor}{Ruzica \surnamestart Piskac\surnameend} \&
  \bibinfo{editor}{Rayna \surnamestart Dimitrova\surnameend}, editors
  (\bibinfo{year}{2016}): \emph{\bibinfo{title}{Proceedings Fifth Workshop on
  Synthesis, SYNT at CAV 2016, Toronto, Canada, July 17-18, 2016}}. {\sl
  \bibinfo{series}{{EPTCS}}} \bibinfo{volume}{229}, \doi{10.4204/EPTCS.229}.

\bibitemdeclare{inproceedings}{PitermanPS06}
\bibitem{PitermanPS06}
\bibinfo{author}{Nir \surnamestart Piterman\surnameend}, \bibinfo{author}{Amir
  \surnamestart Pnueli\surnameend} \& \bibinfo{author}{Yaniv \surnamestart
  Sa'ar\surnameend} (\bibinfo{year}{2006}): \emph{\bibinfo{title}{Synthesis of
  Reactive(1) Designs}}.
\newblock In: {\sl \bibinfo{booktitle}{VMCAI}}, pp. \bibinfo{pages}{364--380},
  \doi{10.1007/11609773\_24}.

\bibitemdeclare{inproceedings}{PR89}
\bibitem{PR89}
\bibinfo{author}{Amir \surnamestart Pnueli\surnameend} \& \bibinfo{author}{Roni
  \surnamestart Rosner\surnameend} (\bibinfo{year}{1989}):
  \emph{\bibinfo{title}{{On the Synthesis of a Reactive Module}}}.
\newblock In: {\sl \bibinfo{booktitle}{POPL}}, \bibinfo{publisher}{{ACM}
  Press}, pp. \bibinfo{pages}{179--190}, \doi{10.1145/75277.75293}.

\bibitemdeclare{inproceedings}{PnueliSZ10}
\bibitem{PnueliSZ10}
\bibinfo{author}{Amir \surnamestart Pnueli\surnameend}, \bibinfo{author}{Yaniv
  \surnamestart Sa'ar\surnameend} \& \bibinfo{author}{Lenore~D. \surnamestart
  Zuck\surnameend} (\bibinfo{year}{2010}): \emph{\bibinfo{title}{{JTLV}: {A}
  Framework for Developing Verification Algorithms}}.
\newblock In: {\sl \bibinfo{booktitle}{CAV}}, {\sl \bibinfo{series}{LNCS}}
  \bibinfo{volume}{6174}, \bibinfo{publisher}{Springer}, pp.
  \bibinfo{pages}{171--174}, \doi{10.1007/978-3-642-14295-6\_18}.

\bibitemdeclare{inproceedings}{RyzhykW16}
\bibitem{RyzhykW16}
\bibinfo{author}{Leonid \surnamestart Ryzhyk\surnameend} \&
  \bibinfo{author}{Adam \surnamestart Walker\surnameend}
  (\bibinfo{year}{2016}): \emph{\bibinfo{title}{Developing a Practical Reactive
  Synthesis Tool: Experience and Lessons Learned}}.
\newblock In \bibinfo{editor}{Piskac} \& \bibinfo{editor}{Dimitrova}
  \cite{DBLP:journals/corr/PiskacD16}, pp. \bibinfo{pages}{84--99},
  \doi{10.4204/EPTCS.229.8}.

\bibitemdeclare{inproceedings}{SchlaipferHB11}
\bibitem{SchlaipferHB11}
\bibinfo{author}{Matthias \surnamestart Schlaipfer\surnameend},
  \bibinfo{author}{Georg \surnamestart Hofferek\surnameend} \&
  \bibinfo{author}{Roderick \surnamestart Bloem\surnameend}
  (\bibinfo{year}{2011}): \emph{\bibinfo{title}{Generalized Reactivity(1)
  Synthesis without a Monolithic Strategy}}.
\newblock In \bibinfo{editor}{Kerstin \surnamestart Eder\surnameend},
  \bibinfo{editor}{Jo{\~{a}}o \surnamestart Louren{\c{c}}o\surnameend} \&
  \bibinfo{editor}{Onn \surnamestart Shehory\surnameend}, editors: {\sl
  \bibinfo{booktitle}{Hardware and Software: Verification and Testing - 7th
  International Haifa Verification Conference, {HVC} 2011, Haifa, Israel,
  December 6-8, 2011, Revised Selected Papers}}, {\sl \bibinfo{series}{Lecture
  Notes in Computer Science}} \bibinfo{volume}{7261},
  \bibinfo{publisher}{Springer}, pp. \bibinfo{pages}{20--34},
  \doi{10.1007/978-3-642-34188-5_6}.

\bibitemdeclare{misc}{CUDD}
\bibitem{CUDD}
\bibinfo{author}{Fabio \surnamestart Somenzi\surnameend}:
  \emph{\bibinfo{title}{{CUDD}: {BDD} package, {U}niversity of {C}olorado,
  {B}oulder.}}
\newblock
  \bibinfo{howpublished}{\url{http://vlsi.colorado.edu/\~fabio/CUDD/cudd.pdf}}.

\bibitemdeclare{inproceedings}{WalkerR14}
\bibitem{WalkerR14}
\bibinfo{author}{Adam \surnamestart Walker\surnameend} \&
  \bibinfo{author}{Leonid \surnamestart Ryzhyk\surnameend}
  (\bibinfo{year}{2014}): \emph{\bibinfo{title}{Predicate abstraction for
  reactive synthesis}}.
\newblock In: {\sl \bibinfo{booktitle}{Formal Methods in Computer-Aided Design,
  {FMCAD} 2014, Lausanne, Switzerland, October 21-24, 2014}},
  \bibinfo{publisher}{{IEEE}}, pp. \bibinfo{pages}{219--226},
  \doi{10.1109/FMCAD.2014.6987617}.

\bibitemdeclare{inproceedings}{YangBOBCJRS98}
\bibitem{YangBOBCJRS98}
\bibinfo{author}{Bwolen \surnamestart Yang\surnameend},
  \bibinfo{author}{Randal~E. \surnamestart Bryant\surnameend},
  \bibinfo{author}{David~R. \surnamestart O'Hallaron\surnameend},
  \bibinfo{author}{Armin \surnamestart Biere\surnameend},
  \bibinfo{author}{Olivier \surnamestart Coudert\surnameend},
  \bibinfo{author}{Geert \surnamestart Janssen\surnameend},
  \bibinfo{author}{Rajeev~K. \surnamestart Ranjan\surnameend} \&
  \bibinfo{author}{Fabio \surnamestart Somenzi\surnameend}
  (\bibinfo{year}{1998}): \emph{\bibinfo{title}{A Performance Study of
  BDD-Based Model Checking}}.
\newblock In \bibinfo{editor}{Ganesh \surnamestart Gopalakrishnan\surnameend}
  \& \bibinfo{editor}{Phillip~J. \surnamestart Windley\surnameend}, editors:
  {\sl \bibinfo{booktitle}{Formal Methods in Computer-Aided Design, Second
  International Conference, {FMCAD} '98, Palo Alto, California, USA, November
  4-6, 1998, Proceedings}}, {\sl \bibinfo{series}{Lecture Notes in Computer
  Science}} \bibinfo{volume}{1522}, \bibinfo{publisher}{Springer}, pp.
  \bibinfo{pages}{255--289}, \doi{10.1007/3-540-49519-3_18}.

\bibitemdeclare{inproceedings}{Zeller99}
\bibitem{Zeller99}
\bibinfo{author}{Andreas \surnamestart Zeller\surnameend}
  (\bibinfo{year}{1999}): \emph{\bibinfo{title}{Yesterday, My Program Worked.
  Today, It Does Not. Why?}}
\newblock In: {\sl \bibinfo{booktitle}{ESEC/FSE}}, {\sl \bibinfo{series}{LNCS}}
  \bibinfo{volume}{1687}, \bibinfo{publisher}{Springer}, pp.
  \bibinfo{pages}{253--267}, \doi{10.1007/3-540-48166-4_16}.

\bibitemdeclare{article}{ZellerH02}
\bibitem{ZellerH02}
\bibinfo{author}{Andreas \surnamestart Zeller\surnameend} \&
  \bibinfo{author}{Ralf \surnamestart Hildebrandt\surnameend}
  (\bibinfo{year}{2002}): \emph{\bibinfo{title}{Simplifying and Isolating
  Failure-Inducing Input}}.
\newblock {\sl \bibinfo{journal}{{IEEE} Trans. Software Eng.}}
  \bibinfo{volume}{28}(\bibinfo{number}{2}), pp. \bibinfo{pages}{183--200},
  \doi{10.1109/32.988498}.

\bibitemdeclare{misc}{optimizationWebsite}
\bibitem{optimizationWebsite}
\emph{\bibinfo{title}{SYNTECH GR(1) Performance Website}}.
\newblock \bibinfo{note}{\url{http://smlab.cs.tau.ac.il/syntech/performance/}}.

\end{thebibliography}

\end{document}